\documentclass[9pt]{article}

\usepackage{txfonts}
\usepackage{times}
\usepackage{url}
\usepackage{booktabs}
\usepackage{multirow}
\usepackage{ccicons}
\usepackage{todonotes}
\usepackage[multiple]{footmisc}
\usepackage{graphicx}
\usepackage{tabularx}
\usepackage{caption}
\usepackage{subcaption}
\usepackage{paralist}
\usepackage{longtable}
\usepackage[utf8]{inputenc}
\usepackage{authblk}

\begin{document}

\title{{Chatty Maps: Constructing sound maps of urban areas from social media data}}



\author[1]{Luca Maria Aiello}
\author[2]{Rossano Schifanella}
\author[3]{Daniele Quercia}
\author[4]{Francesco Aletta}

\affil[1]{Yahoo Labs, London, UK}
\affil[2]{University of Turin, Italy}
\affil[3]{Bell Labs, Cambridge, UK}
\affil[4]{University of Sheffield, UK}

\date{}
\maketitle

\begin{abstract}
Urban sound has a huge influence over how we perceive places. Yet, city planning is concerned mainly with noise, simply because annoying sounds come to the attention of city officials in the form of complaints, while general urban sounds do not come to the attention as they cannot be easily captured at city scale.  To capture both unpleasant and pleasant sounds, we applied a new methodology that relies on tagging information of geo-referenced pictures to the cities of London and Barcelona. To begin with, we compiled the  first urban sound dictionary and compared it to the one produced by collating insights from the literature: ours was experimentally more valid (if correlated with official noise pollution levels) and offered a  wider geographic coverage. From picture tags, we then studied the relationship between soundscapes and emotions. We learned that streets with music sounds were associated with strong emotions of joy or sadness, while those with human sounds were associated with joy or surprise. Finally, we studied the relationship between soundscapes and people's perceptions and, in so doing, we were able to map which areas are chaotic, monotonous, calm, and exciting.Those insights promise to inform the creation of restorative experiences in our increasingly urbanized world. 
\end{abstract}

\section{Introduction}\label{sec:intro}

Studies have found that long-term  exposure to urban noise (in particular, to traffic noise) results into sleeplessness and stress~\cite{halonen15road}, increased incidence of learning impairments among children~\cite{stansfeld05aircraft}, and increased risk of cardiovascular morbidity such as hypertension~\cite{van12quantitative} and heart attacks~\cite{hoffmann06residence,selander09exposure}. 

Since those health hazards are likely to reduce life expectancy, a variety of technologies for  noise monitoring and mitigation have been developed over the years. However, those solutions are costly and do not scale at the level of an entire city. City officials typically measure noise by  placing  sensors at a few selected points. They do so mainly because they have to  comply with  the Environmental Noise Directive (END)~\cite{directive02directive}, which requires the management of noise levels only from specific sources, such as road traffic, railways, major airports and industry. To fix the lack of scalability of a typical solution based on sensors,  in distinct fields, researchers have worked on ways of making noise pollution estimation cheap. They have worked, for example, on epidemiological models to estimate noise levels from a few samples~\cite{morley15international}, on capturing samples from smartphones or other pervasive devices~\cite{maisonneuve09citizen,schweizer11noisemap,meurisch13noisemap,becker13awareness,mydlarz14design}, and on mining geo-located data readily available from social media (e.g., foursquare, twitter)~\cite{hsieh15noisy}. 

All this work has focused, however,  on the negative side of urban sounds. Pleasant sounds have been left out from the urban planning literature, yet they have been shown to  positively impact city dwellers' health~\cite{nilsson06soundscape,andringa13pleasant}. Only a few researchers have been interested in the whole ``urban \textit{soundscape}''. In the World Soundscape Project\footnote{\url{www.sfu.ca/~truax/wsp.html}}, for example, composer Raymond Murray Schafer and colleagues defined soundscape for the first time as \textit{``an environment of sound (or sonic environment) with emphasis on the way it is perceived and understood by the individual, or by a society''}~\cite{schafer77soundscape}. That early work  eventually led to a new International Standard, ISO 12913, where soundscape is defined as \textit{``[the] acoustic environment as perceived or experienced and/or understood by a person or people, in context''}~\cite{iso14soundscape}. Since that work, there remains a number of unsolved challenges though. 

First, there is no shared vocabulary of urban sounds. Back in the early days of the  World Soundscape Project, scholars collected sound-related terms and provided a classification of sounds~\cite{schafer77soundscape}, but that classification was meant to be neither comprehensive  nor systematic. Signal processing techniques for automatically classifying sounds
have recently used labeled examples~\cite{salamon14dataset,salamon15acoustics}, but, again, those training labels are not organized in  any formal taxonomy. 

Second, studying the relationship between urban sounds and people's perceptions is hard. So far the assumption has been that a good proxy for perceptions is noise level. But perceptions depend on a variety of factors; for example, on what one is doing (e.g., whether (s)he is at a concert). Therefore, policies focusing only on the reduction of noise levels might well fall short. 

Finally,  urban sounds cannot be captured at scale and, consequently, they are not considered when planning cities~\cite{aletta14towards}. That is because the collection of data for managing urban acoustic environments has mainly been relegated to  small-scale surveys~\cite{herranz10proposed}~\cite{schulte06recent,schulte13introduction,davies13special}.

To partly address those challenges, we used geo-referenced social media data to  map the soundscape of an entire city, and related that mapping to people's  emotional responses. We did so by extending previous work that captured urban smellscapes from social media~\cite{quercia15smelly} with four main contributions:
\begin{itemize}

\item We collected sound-related terms from different online and offline sources and arranged those terms in a taxonomy.  The taxonomy was determined by matching the sound-related terms with  the tags on 1.8 million geo-referenced Flickr pictures  in Barcelona and London, and by then analyzing how those terms  co-occured  across the pictures to obtain a term classification (co-occuring terms are expected to be semantically related). In so doing, we compiled the first urban sound dictionary and made it publicly available: in it, terms are best classified into 6 top-level categories (i.e., transport, mechanical, human, music, nature, indoor), and those categories closely resemble the manual classification previously derived by aural researchers over decades. 

\item Upon our picture tags, we produced detailed sound maps of Barcelona and London at the level of street segment.	By looking at different segment types, we validate that, as one expects, pedestrian streets host people, music, and indoor sounds, while primary roads are about transport and mechanical sounds. 

\item For the first time, we studied the relationship between urban sounds and emotions. By matching 
	 our picture tags with the terms of a widely-used word-emotion lexicon, we determined people's 
	emotional responses across the city, and how those responses related to urban sound: fear and anger were found on streets with mechanical sounds, while joy was found on streets with human and music sounds. 
	
\item Finally, we studied the relationship between a street's sounds and the perceptions people are likely to have of that street. Perceptions came from  \textit{soundwalks} conducted  in two cities in UK and Italy: locals were asked to identify sound sources and report them along with their subjective perceptions. Then, from social media data, we determined a location's expected perception based on the sound tags at the location. 

\end{itemize}

\section{Methodology}\label{sec:method}

The main idea behind our method was to search for sound-related words (mainly words reflecting potential sources of sound) on geo-referenced social media content. To that end, we needed to get hold of two elements: the sound-related words, and  the content against which to match those words. 

\subsection{Sound Words}\label{sec:method:dictionaries}

We obtained sound-related words from the most comprehensive research project in the field  -- the World Soundscape Project -- and  from  the most popular crowd-sourced online repository of sounds -- Freesound. 

\subsubsection*{Schafer's Words}

The World Soundscape Project is an international research project that initiated the modern study of acoustic ecology.  In his book ``The Soundscape'', the founder of the project, R. Murray Schafer, coined the term soundscape and emphasized the importance of identifying pleasant sounds and using them to create healthier environments. He described how to classify sounds, appreciating their beauty or ugliness, and offered exercises (e.g., ``soundwalks'') to help people become more sensitive to sounds. An entire chapter was dedicated to the classification of urban sounds, which was based on literary, anthropological, and historical documents. Sounds were classified into six categories: natural, human, societal (e.g., domestic sounds), mechanical, quiet, and indicators (e.g., horns and whistles). Our work used that classification as well: to associate words with each category, three annotators independently  hand-coded the book's sections dedicated to the category, and the intersection of the three annotation sets (which is more conservative than the union) was considered,  resulting in a list of 236 English terms.

\subsubsection*{Crowdsourced Words}

Freesound is the largest public online collaborative repository of audio samples: 130K sounds annotated with 1.5M tags  are publicly available through an API. Out of the unique tags (which were 65K), we considered only those that occurred more than 100 times (the remaining ones were too sparse to be useful), resulting in 2.2K tags, which still amounted to 76\% of the total volume as the tag frequency distribution was skewed. However, those tags covered many topics (including usernames, navigational markers, sound quality descriptions, and synthesized sound effects) and reflected ambiguous words at times (e.g., ``fan'' might be a person or a mechanical device) and, as such, needed to be further filtered to retain only words related to sounds or physical sound sources. One annotator manually performed that filtering, which  resulted into a final set of 229 English terms.

In addition to that set of words, there is an online repository specifically tailored to urban sounds called Favouritesounds\footnote{\url{favouritesounds.org}}. This site hosts crowdsourced maps of sounds for several cities in the world: individuals upload recordings of their favorite sounds, place them on the city map,  and annotate them  with free-text descriptions. By manually parsing the 6K unique words contained in those descriptions, we extracted 243 English terms. 

\subsection{Geo-referenced content}\label{sec:method:socialmedia}

Having two sets of sound-related words at hand, we needed social media data against which those words had to be matched. 17M Flickr photos taken between 2005 and 2015 along with their tags were made publicly available in London and Barcelona. In those two cities, we identified each \textit{street segment} from OpenStreetMap\footnote{A segment is often a street's portion between two road intersections but, more generally, it includes any outdoor place (e.g., highways, squares, steps, footpaths, cycle-ways).} (OSM is a global group of volunteers who maintain free crowdsourced online maps). We then collated tags in each segment together by considering the augmented area of the segment's polyline, an area with an extra space of 22.5 meters on each side to account for positioning errors typically present in geo-referenced pictures~\cite{quercia15smelly,quercia15walk}.

\begin{figure}[tp]
\centering
\includegraphics[clip=true, width=.99\textwidth]{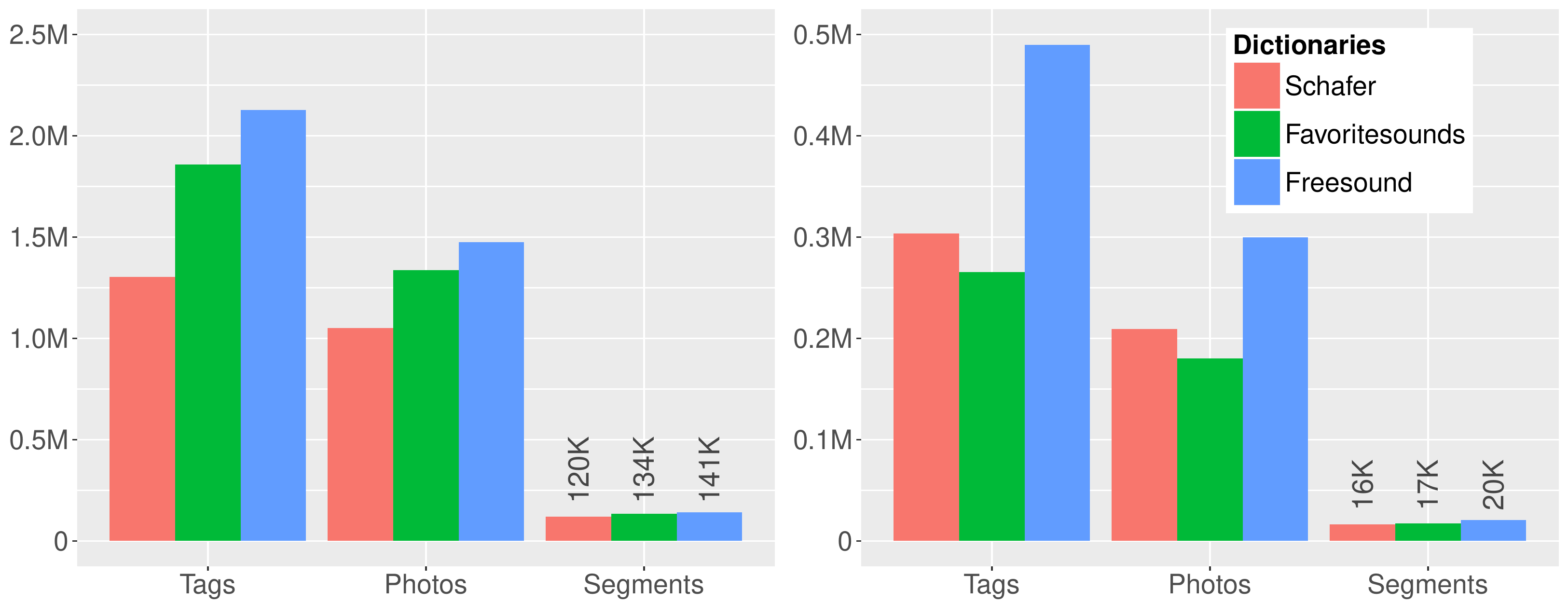}
\caption{Coverage of the three urban sound dictionaries. Number of tags, photos, street segments that had at least one smell word from each vocabulary in Barcelona and London. Each bar is a smell vocabulary. Shafer was extracted from Shafer's book ``The Soundscape'', while the other two were online repositories. The best coverage was offered by FreeSound.}
\label{fig:dictionary_coverage}
\end{figure}

\begin{figure}[tp]
\centering
\includegraphics[clip=true, width=.85\textwidth]{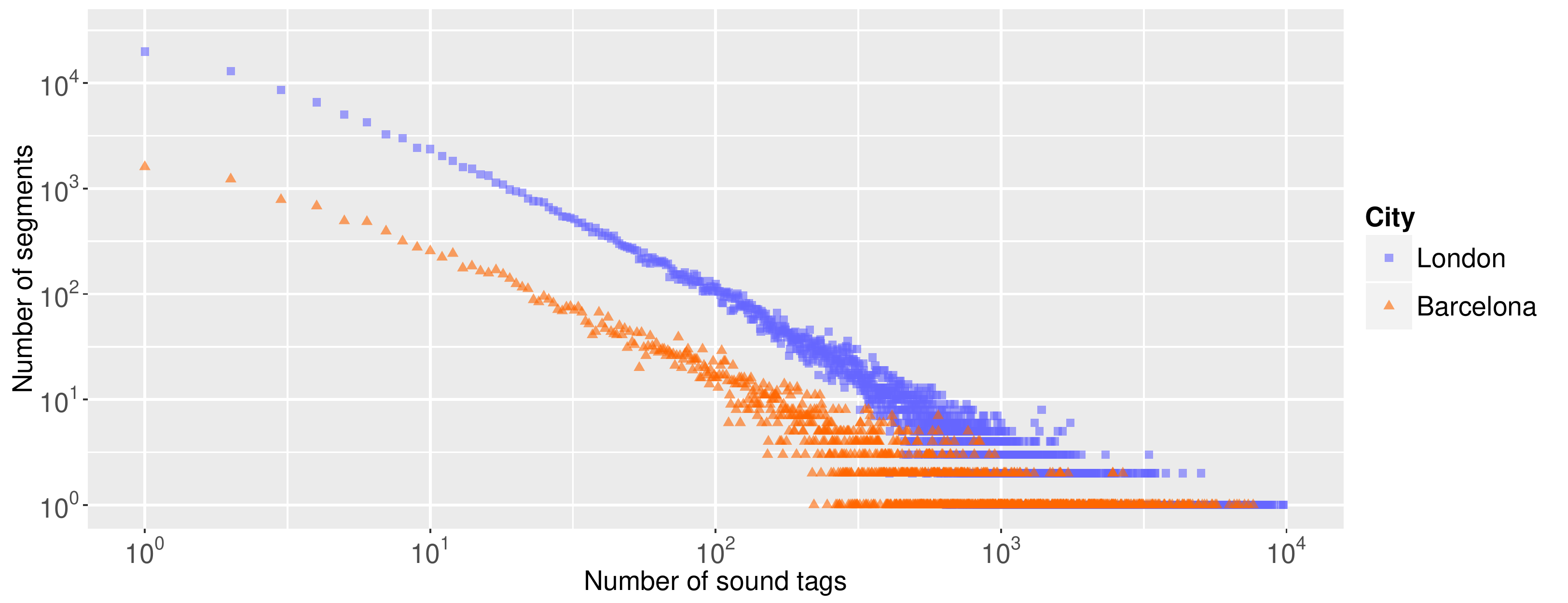}
\caption{Number of street segments ($y$-axis) containing a given number of picture tags that match Freesound terms  ($x$-axis) in London and Barcelona. Many streets had a few tags, and only a few streets have a massive number of them. London has $141K$ segments with at least one tag (and 15 tags in each segment, on average), Barcelona $20K$ (25 tags per segment on average).}
\label{fig:tag_per_segment}
\end{figure}

We found that, among the three crowdsourced repositories, Freesound words matched most of the picture tags and offered the widest geographic coverage (Figure~\ref{fig:dictionary_coverage}), in that, they matched 2.12M tags and covered 141K street segments in London. Also, Freesound's words offered a far better coverage than what Shafer's did, with a broad distribution of tags over street segments (Figure~\ref{fig:tag_per_segment}).

Since the words of the other online repository considerably overlapped with Freesound's ($67\%$ Favoritesounds tags are also in Freesound), we worked only with Freesound (to ensure effective coverage) and with Shafer's classification (to allow for comparability with the literature).

\begin{figure}[tp]
\centering
\includegraphics[clip=true, width=.70\textwidth]{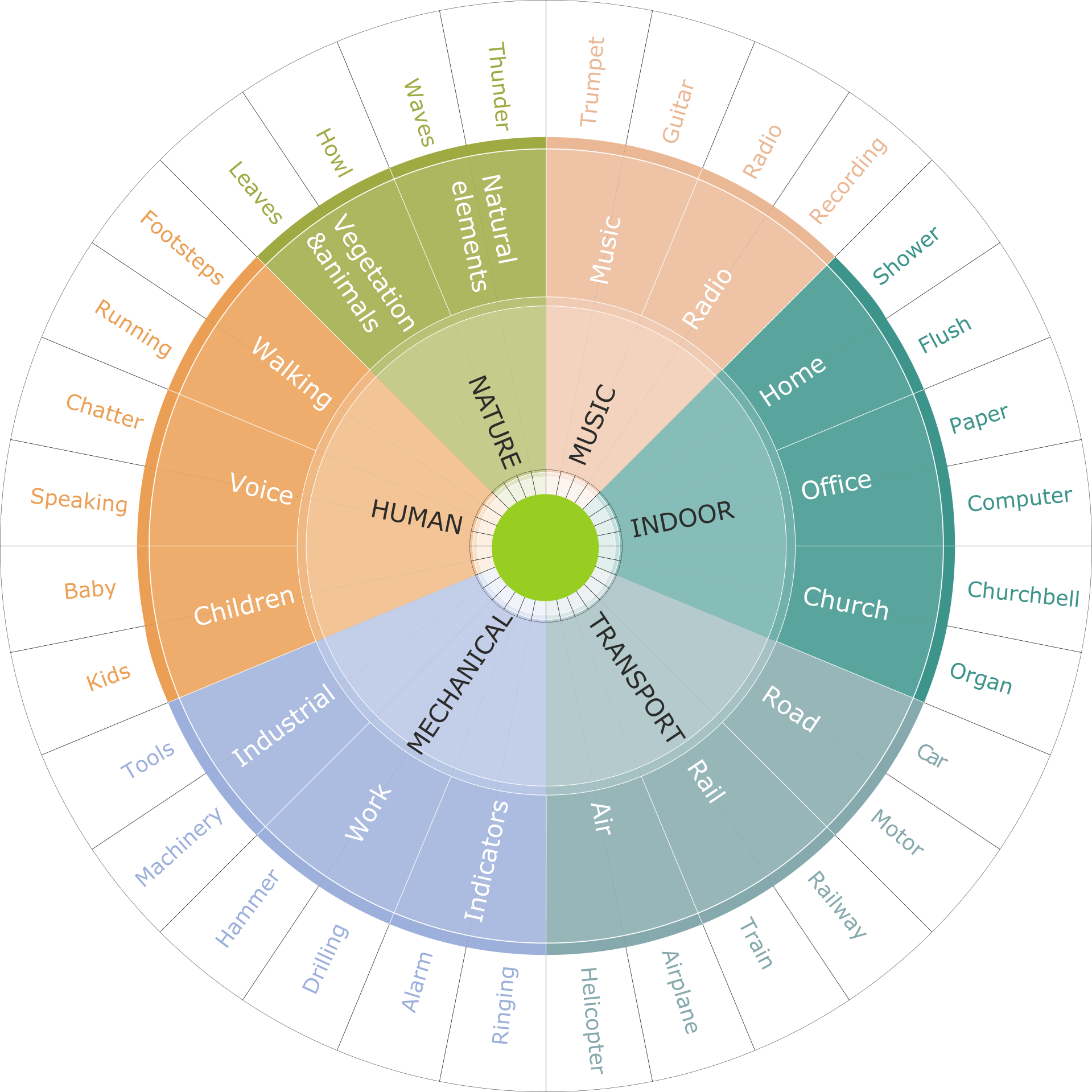}
\caption{Urban sound taxonomy. Top-level categories are in the inner circle; second-level categories are in the outer ring; and examples of words are in the outermost ring. For space limitation, in the wheel, only the first categories (those in the inner circle) are complete, while subcategories and words represent just a sample.}
\label{fig:urban_soundwheel}
\end{figure}

\subsection{Categorization}\label{sec:method:taxonomy}

To discover similarities, contrasts, and patterns, sound words needed to be classified. Schafer already did so. Our Schafer's words are classified into 7 main categories -   \textit{nature}, \textit{human}, \textit{society}, \textit{transport}, \textit{mechanical}, \textit{indicators}, and \textit{quiet} - and each category might have a subcategory (e.g., society includes the subcategories \textit{indoor} and \textit{entertainment}). By contrast, Freesound's words are not classified. However, by looking at which Freesound's words co-occur in the same locations, we could discover similarities (e.g., nature words could co-occur in parks, while transport words in trafficked streets). The use of community detection to extract word categories had been successfully  tested in previous work that extracted categories of smell words~\cite{quercia15smelly}. Compared to other clustering techniques (e.g., LDA~\cite{blei03latent}, K-means~\cite{lloyd82least}), a community detection technique has the advantage of being fully non-parametric and quite resilient to data sparsity.  Therefore, we applied it here as well. We first built a co-occurrence network where nodes  were Freesound's words, and undirected edges were weighted with the number of times the two words co-occurred in the same Flickr pictures as tags. The semantic relatedness among words naturally emerged from the network's community structure: semantically related nodes ended up being both highly clustered together and weakly connected to the rest of the network. To determine the clustering structure, we could have used any of the literally thousands of different community detection algorithms that have been developed in the last decade~\cite{fortunato10community}. None of them always returns the ``best'' clustering. However, since Infomap had shown very good performance across several benchmarks~\cite{fortunato10community}, we opted for it to obtain the initial partition of our network~\cite{rosvall08maps}. Infomap's partitioning resulted in many clusters containing semantically-related words, but it also resulted in some clusters that were simply too big to possibly be semantically homogeneous. To further split those clusters, we applied the community detection algorithm by Blondel \emph{et al.}~\cite{blondel08fast}, which has been found to be the second best performing algorithm~\cite{fortunato10community}. This algorithm stops when no ``node switch'' between communities increases the overall \textit{modularity}~\cite{newman06modularity}, which measures the overall quality of the resulting partitions\footnote{If one were to apply Blondel's right from the start, the resulting clusters would be less coherent than those produced by our approach.}. The result of those two steps is the grouping of sound words in hierarchical categories. Since a few partitions of words could have been too fine-grained, we manually double-checked whether this was the case and, if so, we merged all those sub-communities that were under the same hierarchical partition and that contained strongly-related sound words.

Figure~\ref{fig:urban_soundwheel} sketches the resulting classification in the form of a sound wheel. This wheel has six main categories (inner circle), each of which has a hierarchical structure with variable depth from 0 to 3. For brevity, the wheel reports only the first level fully (inner circle), while it reports samples for the two other levels. Despite spontaneously emerging from word co-occurrences and being fully data-driven, the classification in the wheel strikingly resembles Schafer's. The three categories \textit{human}, \textit{nature}, and \textit{transport} are all in both categorizations. The category \textit{quiet} is missing because it does not match any tag in Freesound, as one would expect. The remaining categories are all present but arranged at a different level: \textit{music} and \textit{indoor} are at the first level in the wheel, while they are at the second level in Schafer's categorization; the  \textit{mechanical} category in the wheel collates two of  Schafer's categories into one: \textit{mechanical} and \textit{indicator}. 

\begin{figure}[tp]
\centering
\includegraphics[clip=true, width=.99\textwidth]{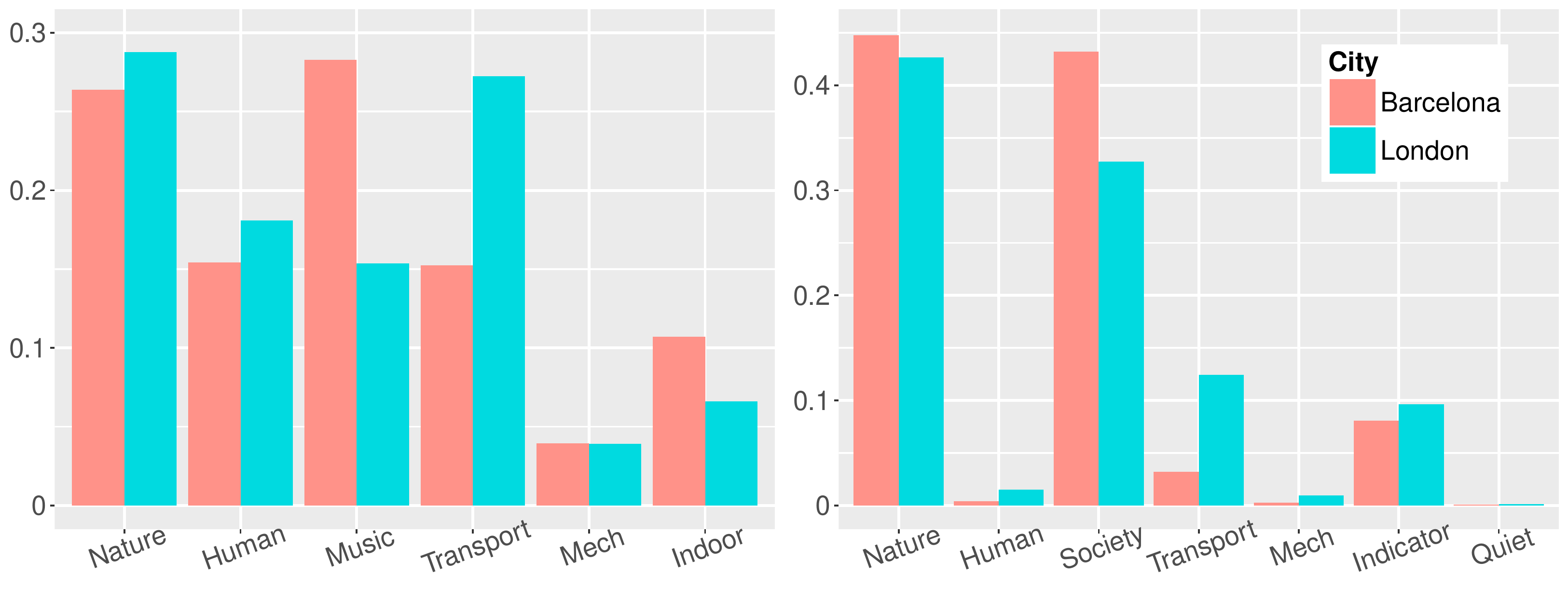}
\caption{Fraction $\frac{tag_c}{tag}$ of picture tags that matched sound category $c$ over all the tags in the city.}
\label{fig:tag_categories_comparison}
\end{figure}

Freesound not only offered a classification similar to Schafer's and to recent working groups' classifications~\cite{brown11towards,salamon14dataset} (speaking to its external validity) but also offered a richer vocabulary of words. By looking at the fraction $\frac{tag_c}{tag}$ of sound words in category $c$ that matched at least one geo-reference picture tag ($tag_c$) over  the total number of tags in the city ($tag$), we saw that Freesound resulted in a full representation of all sound categories  (Figure~\ref{fig:tag_categories_comparison}), while  Shafer's resulted in a patchy representation of many categories.  Therefore, given its effectiveness, Freesound was chosen as the sound vocabulary for the creation of the urban sound wheel. Only the wheel's top-level categories were used. The full taxonomy is, however, available online\footnote{\url{http://goodcitylife.org/chattymaps/project.html}} for those wishing to explore specialized aspects of urban sounds (e.g., transport, nature). 

\section{Validation}\label{sec:results}

With our sound categorization, we were able to determine, for each street segment $j$, its \textit{sound profile} $sound_{j}$ in the form of a 6-element vector. Given sound category $c$, the element $sound_{j,c}$ is:
\begin{equation}
sound_{j,c} = \frac{tag_{j,c}}{tag_{j}};
\label{eq:sound_profiles}
\end{equation}
where $tag_{j,c}$ is the number of tags  at segment $j$ that matched sound category $c$,  and $tag_{j}$ is the total number of tags at segment $j$. To make sure the sound categories $c$'s we had chosen resulted into reasonable outcomes, we verified whether different street types were associated with sound profiles one would expect (Section~\ref{sec:results:soundprofile}), and whether those profiles matched official noise pollution data  (Section~\ref{sec:results:noise}). 

\subsection{Street types}\label{sec:results:soundprofile}
\begin{figure}[tp]
\centering
\includegraphics[clip=true, width=.70\textwidth]{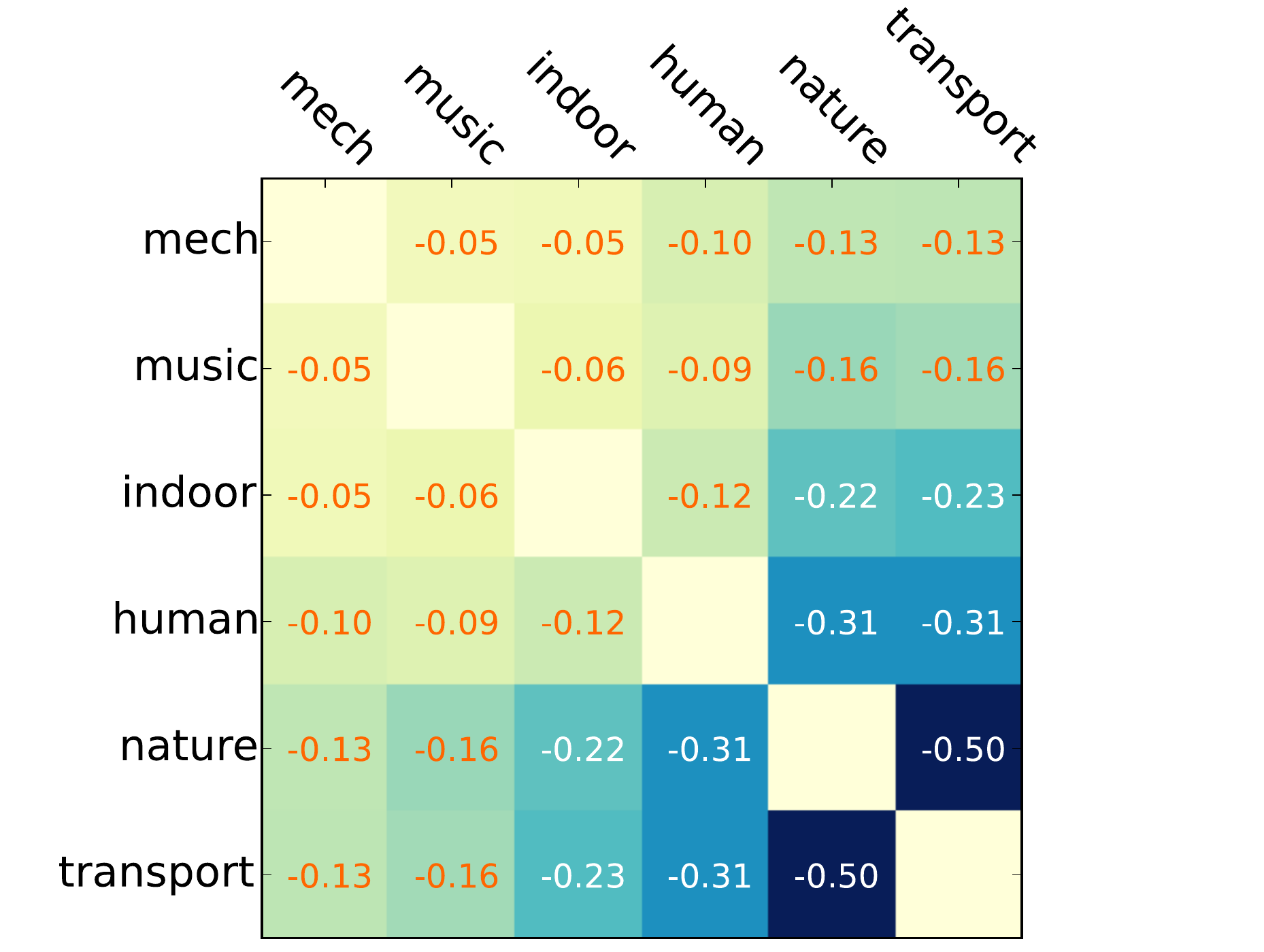}
\caption{Pairwise rank correlations between the fraction of sound tags in category $c_1$ ($sound_{j,c_1}$) and the faction in category $c_2$ ($sound_{j,c_2}$) across all segments $j $'s in London.}
\label{fig:cross_correlation}
\end{figure}

One way of testing whether the 6-category classification makes sense in  the city context is to see which pairs of categories do not tend to co-occur spatially (e.g., nature and transport should be on separate streets). Therefore, for each street segment, we computed the pairwise Spearman rank correlation $\rho$ between  the fraction of sound tags in category $c_1$ and that of sound tags in category $c_2$, across all segments (Figure~\ref{fig:cross_correlation}).  That is, we computed  $\rho_j(sound_{j,c_1}, sound_{j,c_2})$ across all $j$'s. We found that the correlations were either zero or negative. This meant that the categories were either orthogonal (i.e., the categories of \textit{human}, \textit{indoor}, \textit{music}, \textit{mechanical} show correlations close to zero) or geographically sorted in expected ways (with a $\rho=-0.50$,  \textit{nature} and \textit{transport} are  seen, on average, on distinct segments).

\begin{figure}[tp]
\centering
\includegraphics[clip=true, width=.90\textwidth]{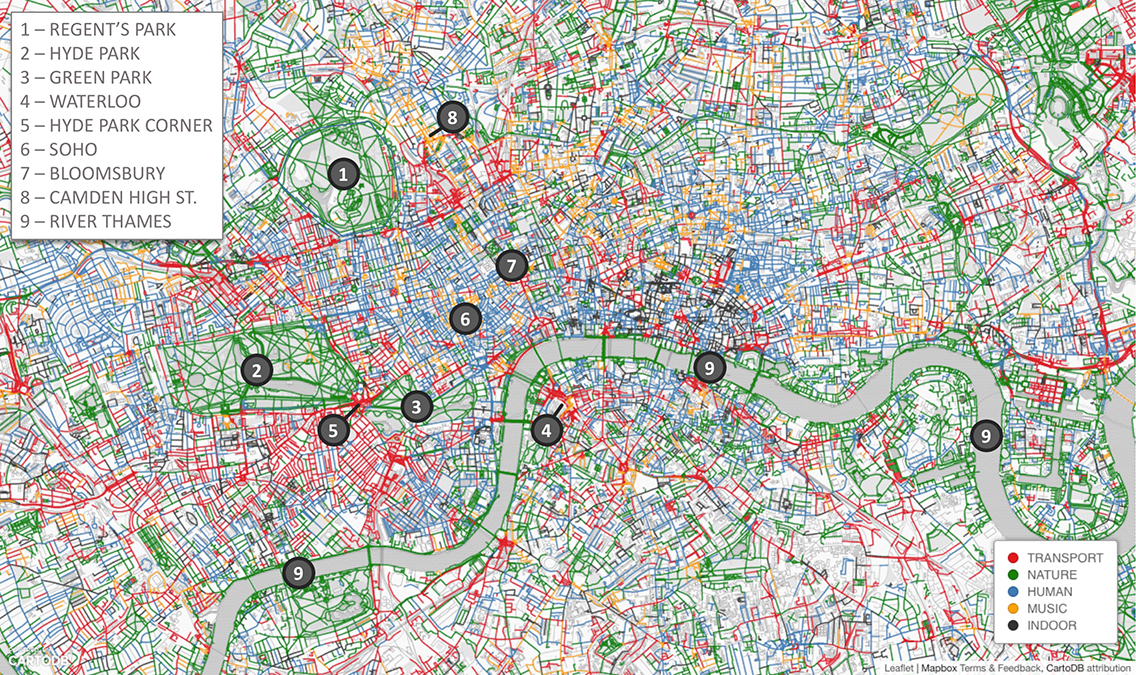}\\
\includegraphics[clip=true, width=.90\textwidth]{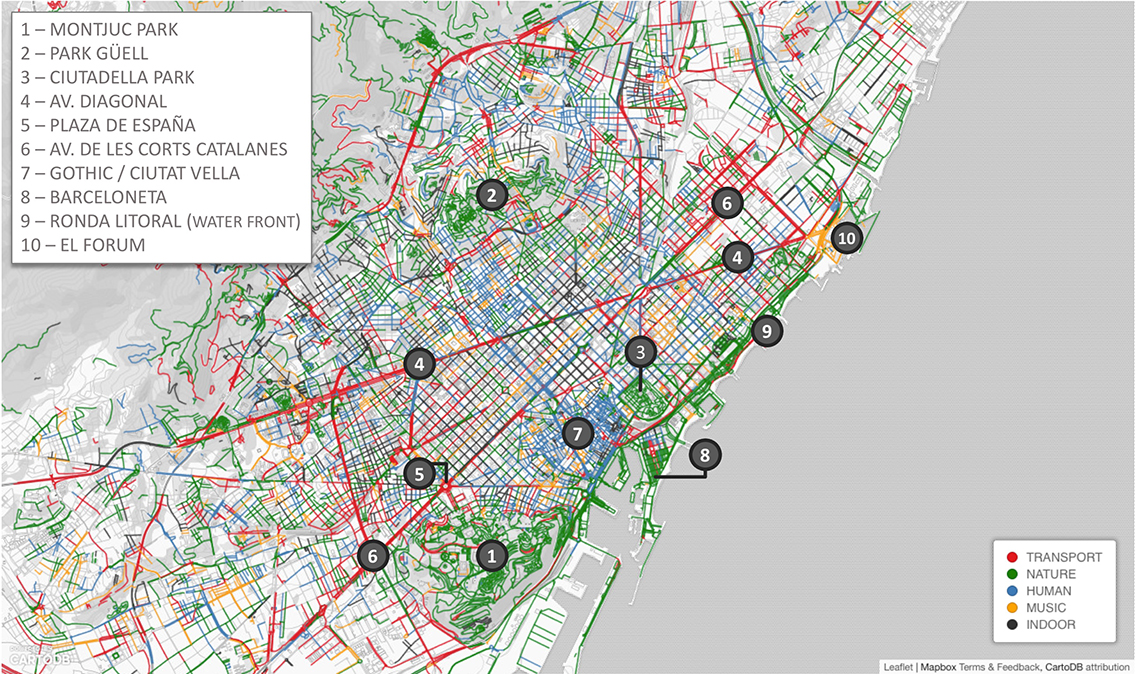}
\caption{Urban Sound Maps of London (top) and Barcelona (bottom). Each street segment is marked with the sound category $c$ that has the highest $z$-score for that segment ($z_{sound_{j,c}}$). In London, natural sounds are found in  Regent's Park(1); Hyde Park(2);  Green Park(3); and all around the  River Thames(9). By contrast, transport sounds are around Waterloo station(4) and on the perimeter of Hyde Park (5). Human sounds are found in Soho(6) and Bloomsbury(7), and music is associated with the small clubs on Camden High Street(8). In Barcelona, natural sounds are found in  Montjuic Park(1); Park Guell(2); and  Ciutadella Park(3), and on the beaches of Barceloneta(8)  and  Ronda Litoral(9). By contrast, annoying and chaotic sounds are found on the main road of  Avinguda Diagonal(4), on Plaza de Espana(5), and on Avinguda De Les Corts Catalanes(6). Human sounds are found in the historical center called Gothic/Ciutat Vella(7), and music in the open-air arena of El Forum(10). Only segments with at least 5 sound tags were considered.}
\label{fig:soundscape_map}
\end{figure}

To visualize the geographic sorting of sounds, we marked each street segment with the sound category that had the highest $z$-score in that segment (Figure~\ref{fig:soundscape_map}). The $z$-scores reflect the extent to which the fraction of sound tags in category $c$ at street segment $j$ deviated from the average fraction of sound tags in $c$ at all the other segments:
\begin{equation}
z_{sound_{j,c}} = \frac{sound_{j,c} - \mu(sound_{c})}{\sigma(sound_{c})};
\label{eq:sound_profile_zscores}
\end{equation}
where $\mu(sound_{c})$ and $\sigma(sound_{c})$ are the mean and standard deviation of the fractions of tags in sound category $c$ across all segments. We then reported  the most prominent sound at each street segment in Figure~\ref{fig:soundscape_map}: \textit{traffic} was associated with street junctions and main roads, \textit{nature} with parks or greenery spots, and \textit{human} and \textit{music} with central parts or with pedestrian streets.

\begin{figure}[tp]
\centering
\includegraphics[clip=true, width=.99\textwidth]{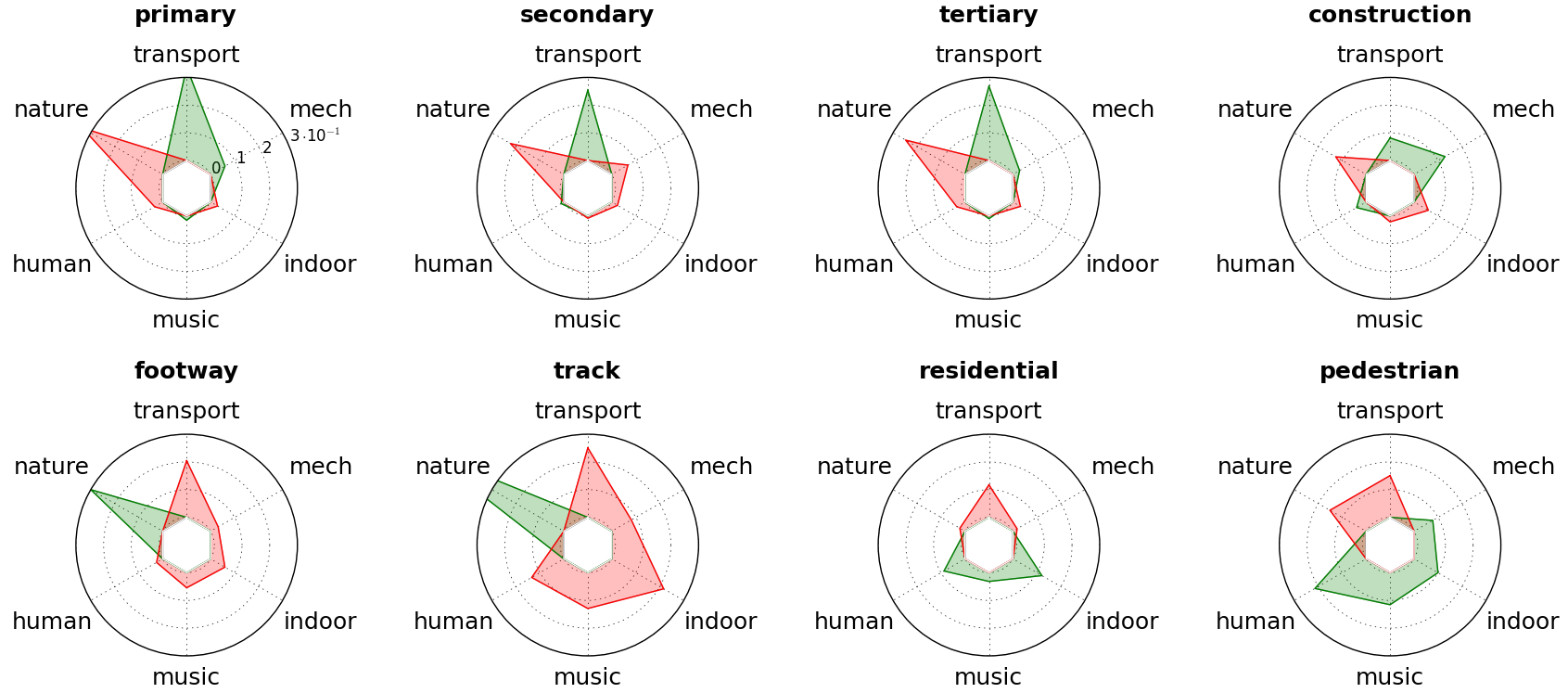}
\caption{Average $z$-scores of the presence of six sound categories for segments of each street type ($\overline{z}_{sound_{c},type_{t}}$). Positive values are in green, and negative ones are in red.  The first clock-like representation refers to primary roads and shows that  transport sounds have z-score of 0.3 (they are present more than average), while  nature sounds  have z-score of -0.3 (they are present less than average). The number of segment per type ranges between 1K and 25K, with the only exception of the ``construction'' type that has only 83 segments. Confidence intervals around the average values range between $10^{-2}$ and $10^{-3}$.}
\label{fig:waytype}
\end{figure}

One indeed expects that different street types (Table~\ref{tab:osmtypes} reports the most frequent types in OSM) would be associated with different sounds. To verify that, we computed the average $z$-score of a sound category $c$ for the segments with street type $t$:

\begin{equation}
\overline{z}_{sound_{c},type_{t}} = \frac{\sum_{j \in S_t} (z_{sound_{j,c})}}{|S_t|};
\label{eq:sound_profile_zscores_average}
\end{equation}
where $S_t$ is the set of segments of (street) type $t$. Figure~\ref{fig:waytype} reports the average values of those $z$-scores. Each clock-like representation refers to a street type, and the sound categories unfold along the clock: positive (negative) $z$-score values are marked in green (red) and suggest a presence of a sound category higher (lower) than the average one. By looking at the positive values , we saw that primary, secondary and tertiary streets (which contain cars) were associated with transport sounds; construction sites with mechanical sounds; footways and tracks (often embedded in parks) were associated with nature sounds; residential  and pedestrian streets were associated with human, music, and indoor sounds. Then, by looking at the negative values, we learned that primary, secondary, tertiary, and construction streets were not associated with nature; and the other street types were not associated with sounds related to transport. 

\begin{table}[tp]
\footnotesize
\begin{tabularx}{390pt}{cX}
\textbf{Street type} & \multicolumn{1}{c}{\textbf{Description}} \\
\hline
Footway       & Designated footpaths mainly or exclusively for pedestrians. This includes walking tracks and gravel paths. \\
Residential   & Roads that serve as an access to housing, without function of connecting settlements. Often lined with housing.  \\
Pedestrian    & Roads used mainly or exclusively for pedestrians in shopping and residential areas. They may allow access of motorised vehicles only for very limited periods of the day. \\
Track         & Roads for mostly agricultural or forestry uses. Tracks are often rough with unpaved surfaces. \\
Primary       & A major highway linking large towns, normally with 2 lanes not separated by a central barrier. \\
Secondary     & A highway which is not part of a major route, but nevertheless forming a link in the national route network, normally with 2 lanes.\\
Tertiary      & Roads connecting smaller settlements or  roads connecting minor streets to more major roads. \\
Construction  & Active road construction sites. Major road and rail construction schemes that typically require several years to complete.\\
\hline
\end{tabularx}
\caption{Description of the eight most frequent street types in Open Street Map.}
\label{tab:osmtypes}
\end{table}

\subsection{Noise pollution}\label{sec:results:noise}
\begin{figure}[tp]
\centering
\includegraphics[clip=true, width=.75\textwidth]{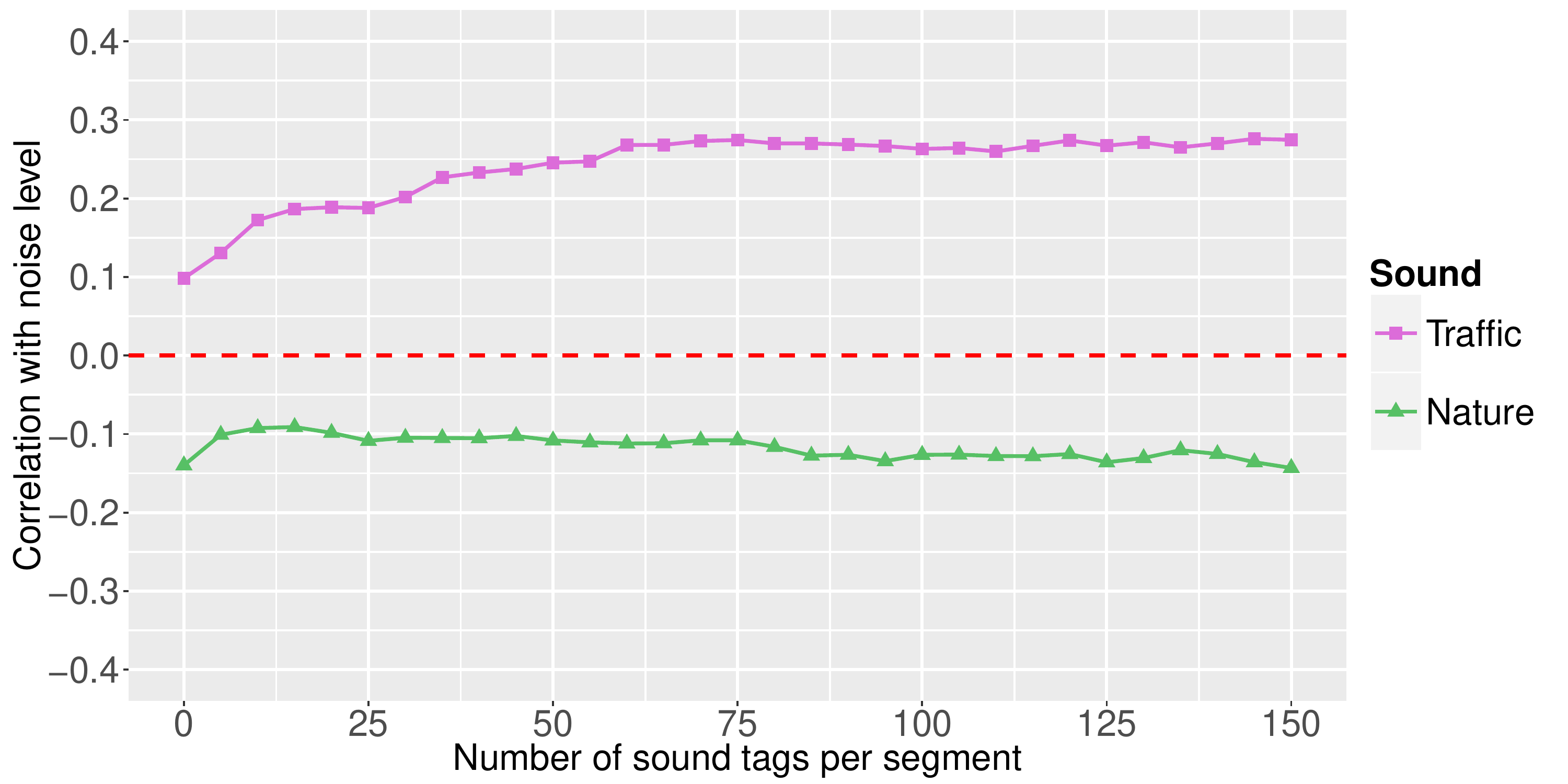}
\caption{Spearman correlation between the fraction of tags at segment $j$ that matched category $c$ ($sound_{j,c}$) and $j$'s noise levels (expressed as Equivalent Weighted Level values in $dB$) as the number of tags per street segment ($x$-axis) increases. All correlations are statistically significant at the level of $p < 0.01$.}
\label{fig:correlation_sound_pollution}
\end{figure}

The most studied aspect of urban sounds is the issue of noise pollution. Despite the importance of that issue, there is no reliable and high-coverage noise measurement data for world-class cities. There is a great number of participatory sensing applications that manage databases of noise levels in several cities, and  some of them are publicly accessible~\cite{maisonneuve09citizen,schweizer11noisemap,meurisch13noisemap,becker13awareness,mydlarz14design}, but all of them offer a limited geographic coverage of a city. 

Barcelona is an exception, however. In 2009, the city council started a project, called Strategic Noise Map, whose main goal was to monitor noise levels and ultimately find new ways of limiting sound pollution. The project has a public API\footnote{Interactive Map of Noise Pollution in Barcelona \url{http://w20.bcn.cat:1100/WebMapaAcustic/mapa_soroll.aspx?lang=en}} that returns noise values at the level of street segment  for the whole city. For each segment, we collected the four $dB$ values provided: three yearly averages for the three times of the day (day, from 7am to 9pm; evening, from 9pm to 11pm; and night, from 11pm to 7am), and one aggregate value, the Equivalent Weighted Level (EWL), that averages those three values adding a 5dB penalty to the evening period, and a 10dB to the night period.   With a practice akin to the one used for air quality indicators~\cite{eeftens12development,beelen13development}, those noise level values are estimated by a prediction model  that is bootstrapped with field measurements~\cite{gulliver15development}. In the case of Barcelona,  the model is bootstrapped with 2.300 short-span noise measurements lasting at most 15 minutes, usually taken during daytime, and with 100 long-span ones lasting from 24 hours to a few days.

To see whether noise pollution was associated with specific sound categories, we considered the street segments with at least $N$ tags and computed, across all the segments, the Spearman rank correlations $\rho_j(EWL_j,sound_{j,c})$ between segment $j$'s EWL values  (in dB) and  $j$'s fraction of picture tags that matched category $c$\footnote{In computing the correlations,  we used the method by Clifford \emph{et al.}~\cite{clifford89assessing} to addresses spatial auto-correlations.} (Figure~\ref{fig:correlation_sound_pollution}). The idea was to determine not only which categories were associated with noise pollution but also how many tags were needed to have a significant association. We found that noise pollution was positively correlated  ($p<0.01$) with \textit{traffic} ($0.1<\rho<0.3$) and negatively correlated with \emph{nature} ($-0.1<\rho<-0.2$), and those results did hold for low values of $N$, suggesting that only a few hundred  tags were needed to build a representative sound profile of a street.

\section{Emotional and Perceptual Layers}\label{sec:other-maps}
Sounds can be classified in ways that reflect aspects other than semantics - they may be classified according to, for example, their emotional qualities or the way they are perceived. Therefore, we now show how social media helps extracting the emotional layer (Section~\ref{sec:results:emotion}) and the perceptual layer (Section~\ref{sec:results:perception}) of urban sound.

\subsection{Emotional Layer}\label{sec:results:emotion}

Looking at a location through the lens of social media makes it possible to characterize places from different points of views. Sound has a highly celebrated link with emotions, especially music sound~\cite{kivy89sound,zentner08emotions}, and it has a considerable effect on our feelings and our behavior. 

One way of extracting emotions from geo-referenced content is to use a word-emotion lexicon known as EmoLex~\cite{mohammad13crowdsourcing}. This lexicon classifies words into eight primary emotions: it contains binary associations of 6,468 terms with the their typical emotional responses. The 8 primary emotions (anger, fear, anticipation, trust, surprise, sadness, joy, and disgust) come from Plutchik's psychoevolutionary theory~\cite{plutchik91emotions}, which is commonly used to characterize general emotional responses. We opted for EmoLex instead of other commonly used sentiment dictionaries (such as LIWC~\cite{pennebaker2013secret}) as it made it possible to study finer-grained emotions. 

\begin{figure}[tp]
\centering
\includegraphics[clip=true, width=.90\textwidth]{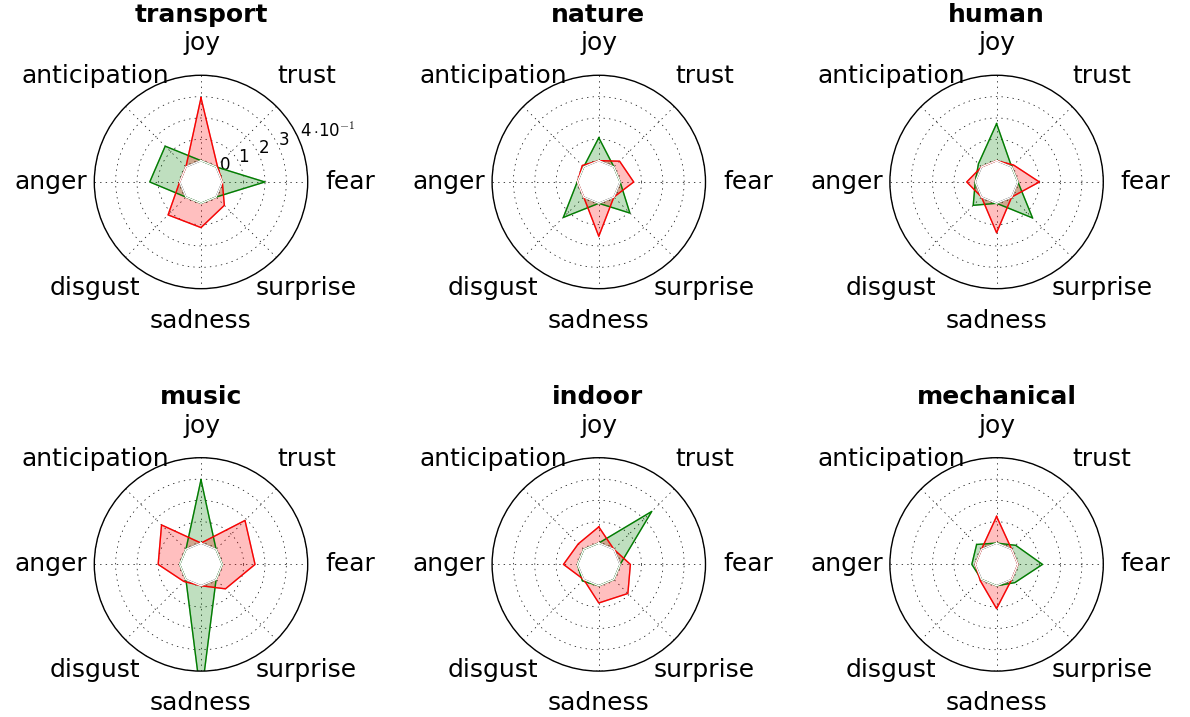}
\caption{Correlation between $z_{sound_{j,c}}$ and  $z_{emotion_{j,e}}$. Each clock-like representation refers to a sound category. The different emotions unfold around the clock, and the emotions that are associated with the sound category are marked in green (positive correlations) or in red (negative emotions). All correlations are statistically significant at the level of $p < 0.01$.}
\label{fig:emolex_sound}
\end{figure}

We matched our Flickr tags with the words in EmoLex  and, for each street segment, we computed its \textit{emotion profile}. The profile consisted of all Plutchik's primary emotions, in that, each of its elements was associated with an emotion:
\begin{equation}
emotion_{j,e} = \frac{tag_{j,e}}{tag_{j}};
\label{eq:emotional_profile}
\end{equation}
where $tag_{j,e}$ is the number of tags at segment $j$ that matched primary emotion $e$. We then computed the corresponding $z$-score:
\begin{equation}
z_{emotion_{j,e}} = \frac{emotion_{j,e} - \mu(emotion_{e})}{\sigma(emotion_{e})}.
\label{eq:emotional_profile_zscores}
\end{equation}
By computing the Spearman rank correlation $\rho_j(z_{sound_{j,c}},z_{emotion_{j,e}})$, we determined which sound was associated with which emotion. From Figure~\ref{fig:emolex_sound}, we see that joyful words were associated with streets typically characterized by music and human sounds, while they were absent in streets with traffic. Traffic was, instead, associated with words of fear, anticipation, and anger. Interestingly, words of sadness (together with those of joy) were associated with streets with music, words of trust with indoors, and words of surprise with streets typically  characterized by human sounds.

\subsection{Perceptual Layer}\label{sec:results:perception}

\begin{figure}[t!]
\includegraphics[width=0.19\columnwidth]{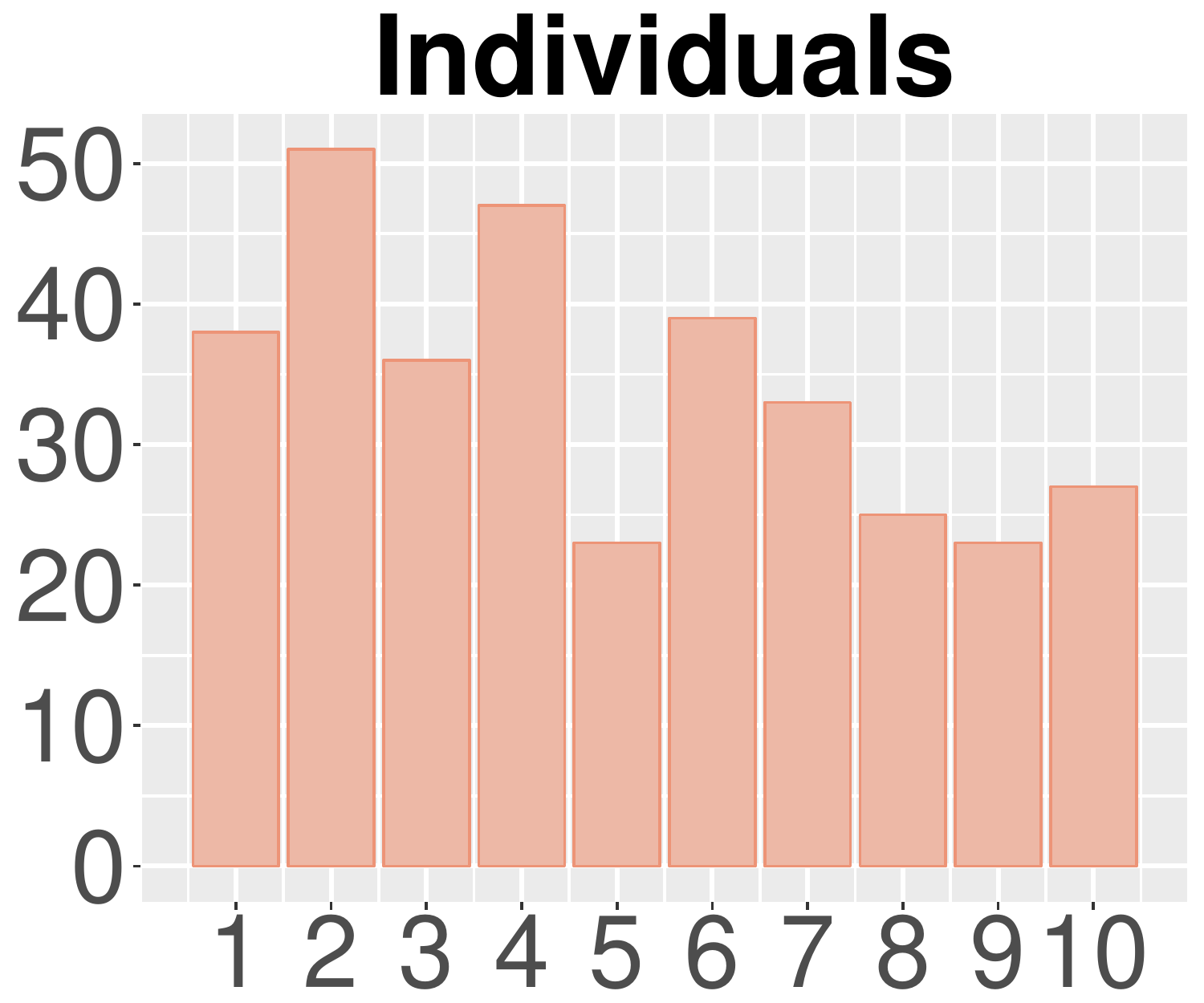} 
\includegraphics[width=0.19\columnwidth]{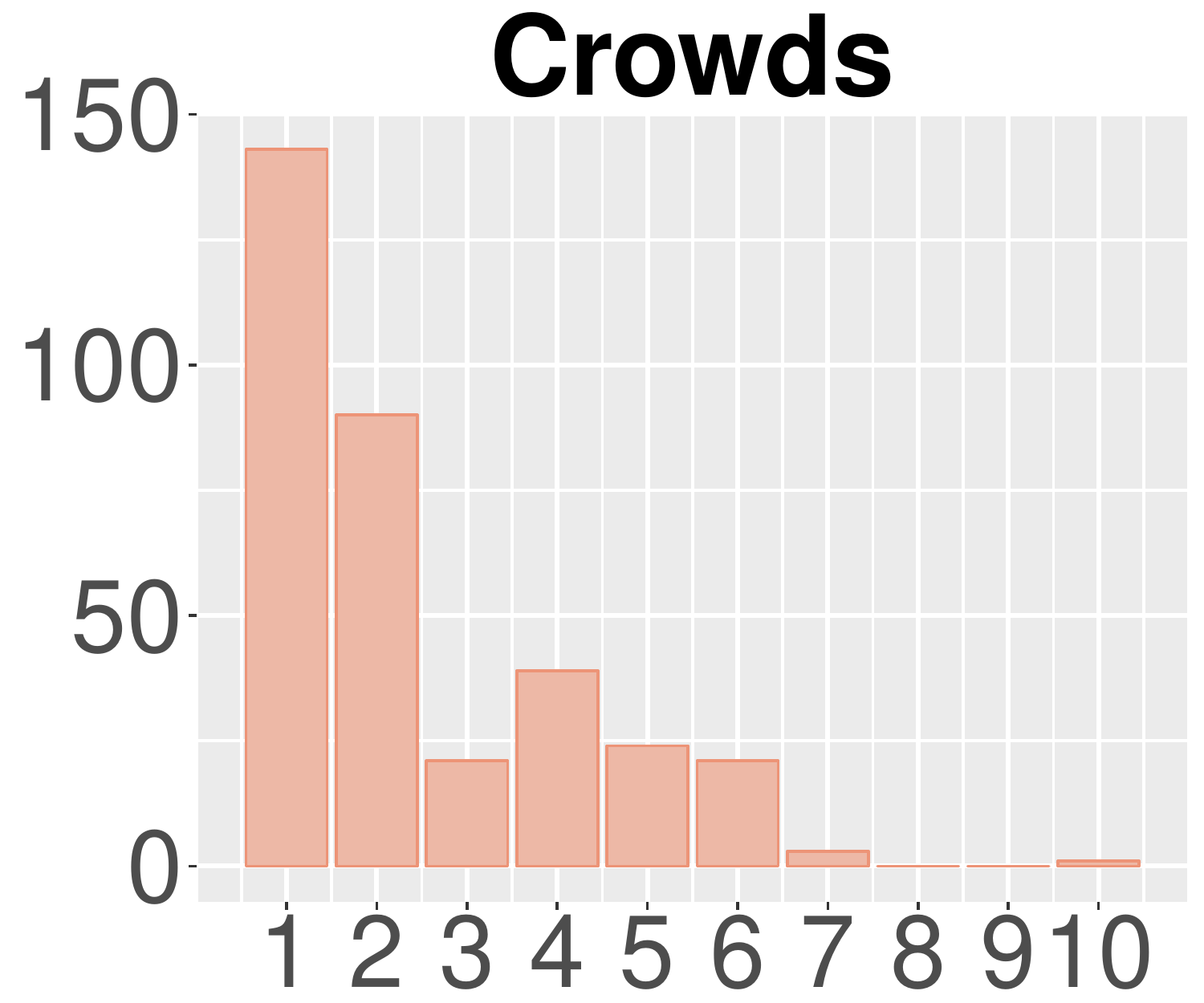}
\includegraphics[width=0.19\columnwidth]{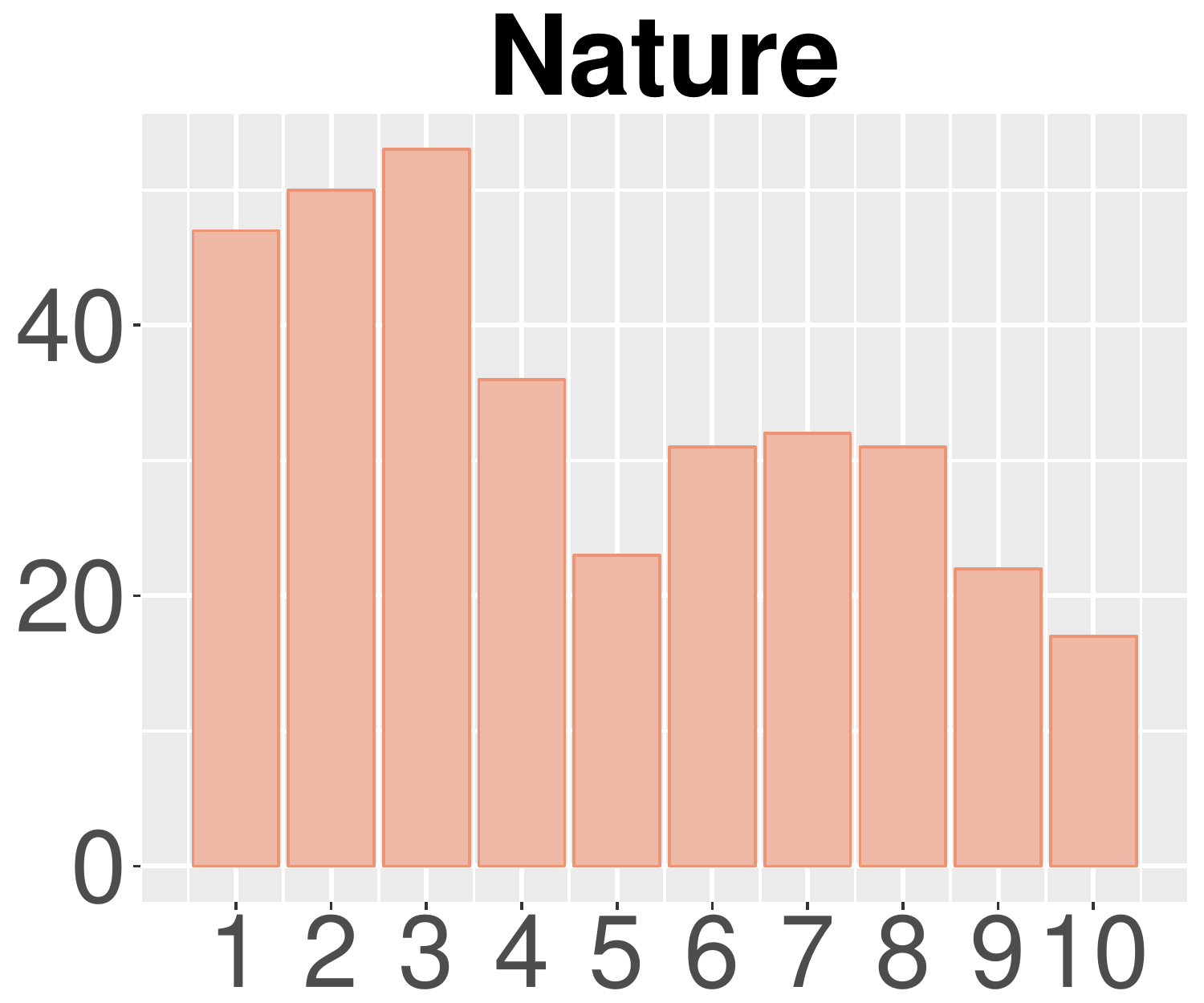}
\includegraphics[width=0.19\columnwidth]{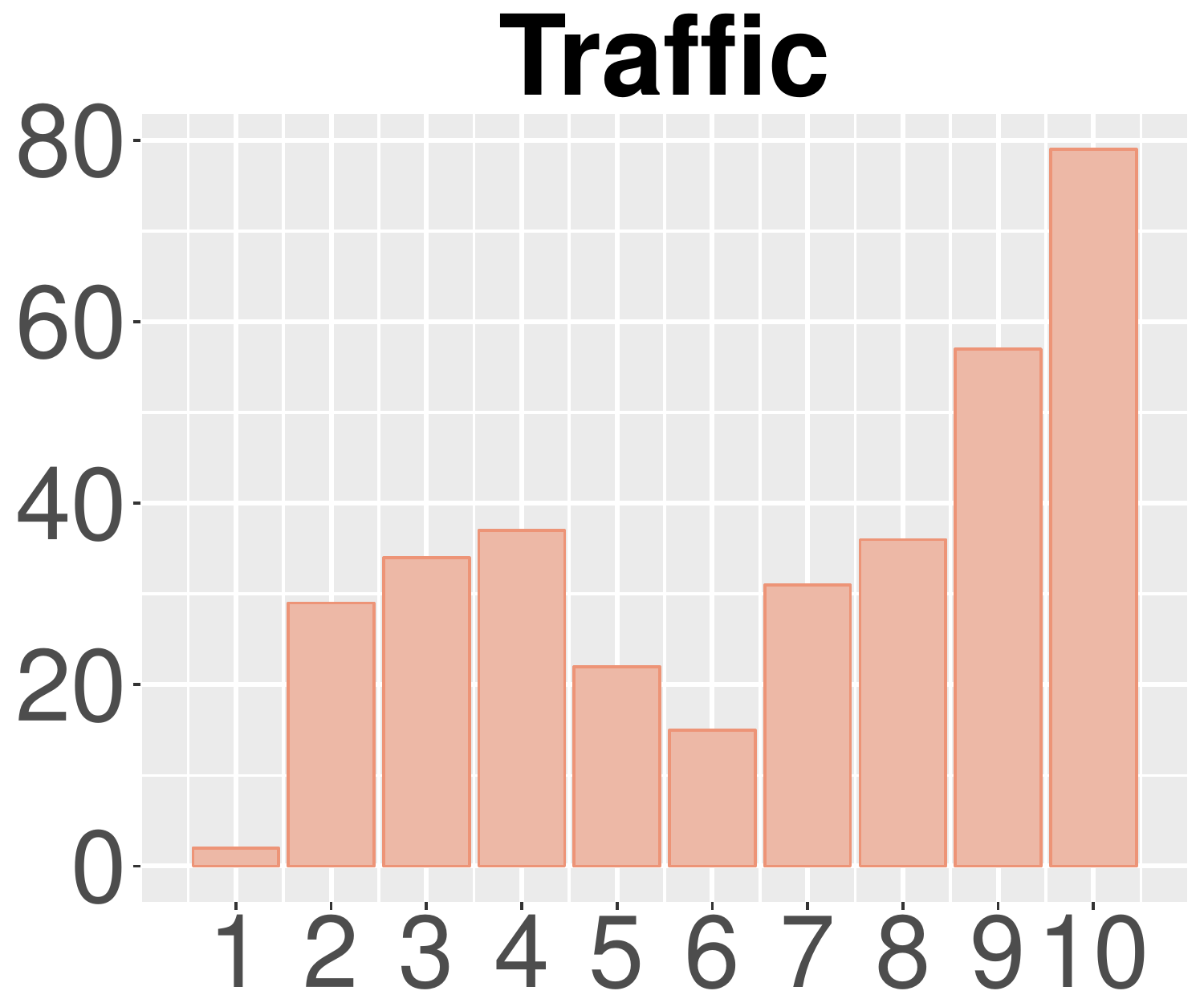}
\includegraphics[width=0.19\columnwidth]{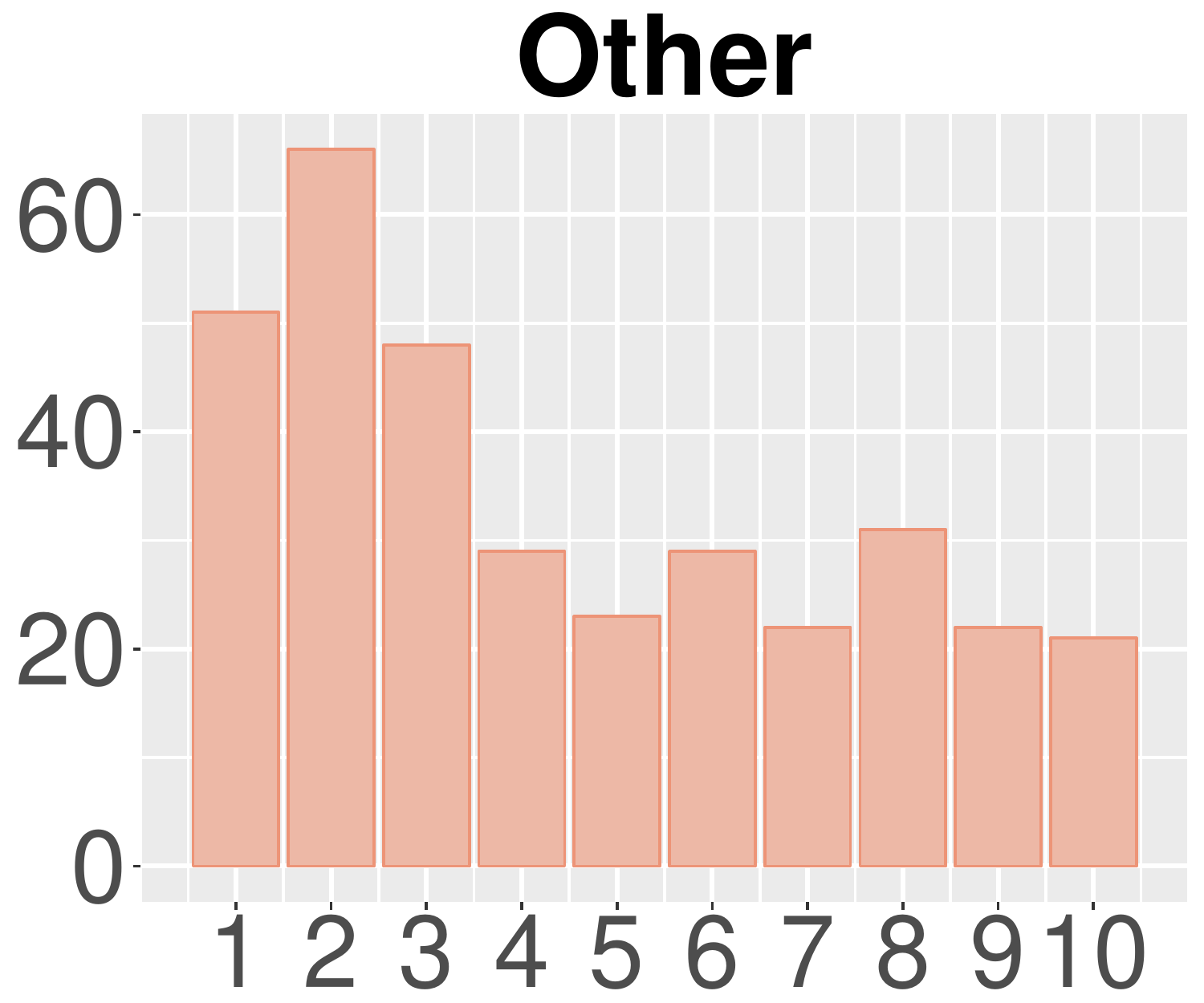}
\caption{Frequency distributions of the survey's scores for sound presence (from 1 to 10) across categories: individuals, crowds, nature, traffic, and other. Sounds of  individuals are scored in  the full 1-to-10 range, while sounds of crowds are typically scored with a value of 1 or 2 as they might have been absent most of the times.}
\label{fig:perception_soundcategories_distr} 
\end{figure}

\begin{figure}[t!]
\includegraphics[width=0.24\columnwidth]{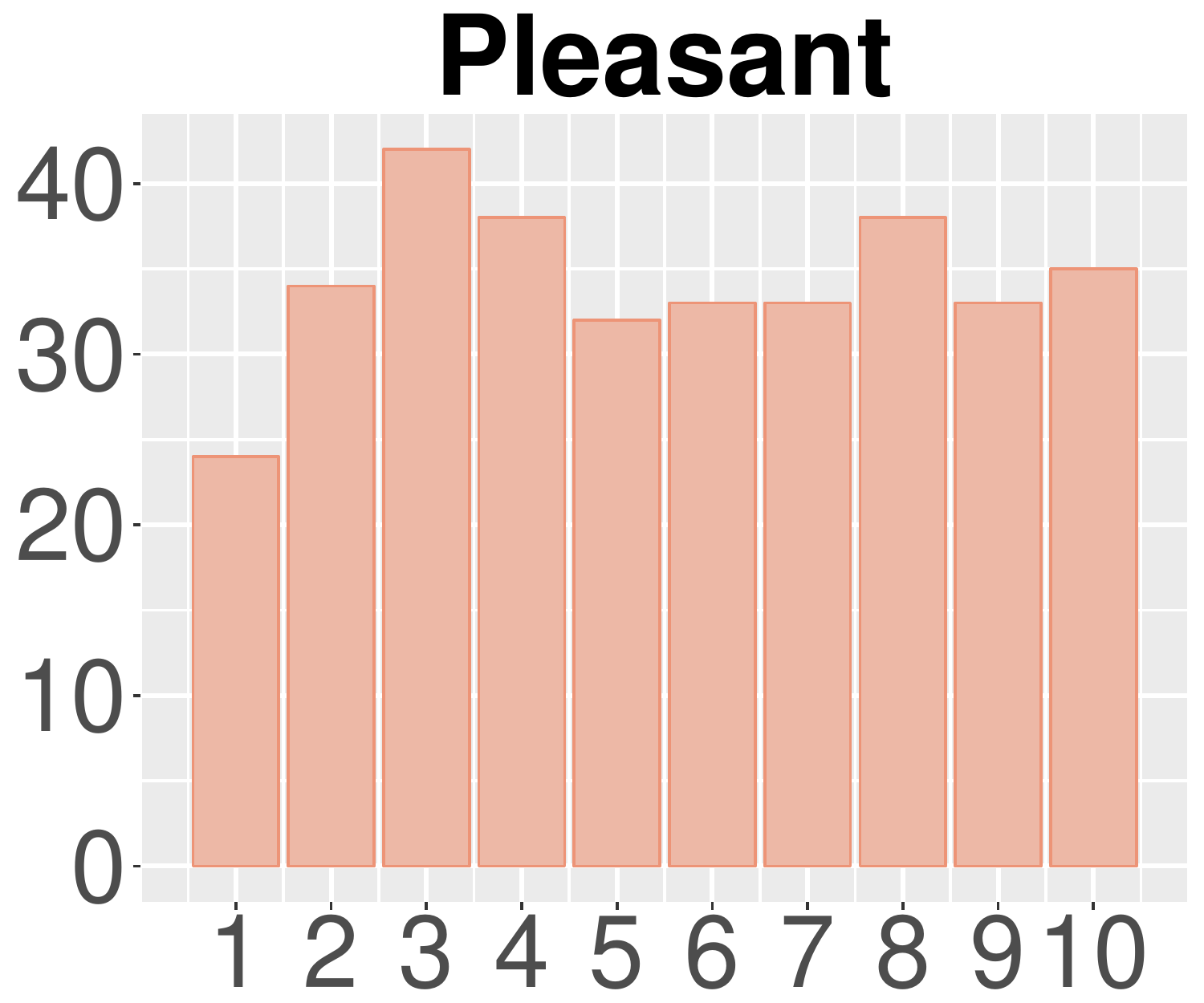} 
\includegraphics[width=0.24\columnwidth]{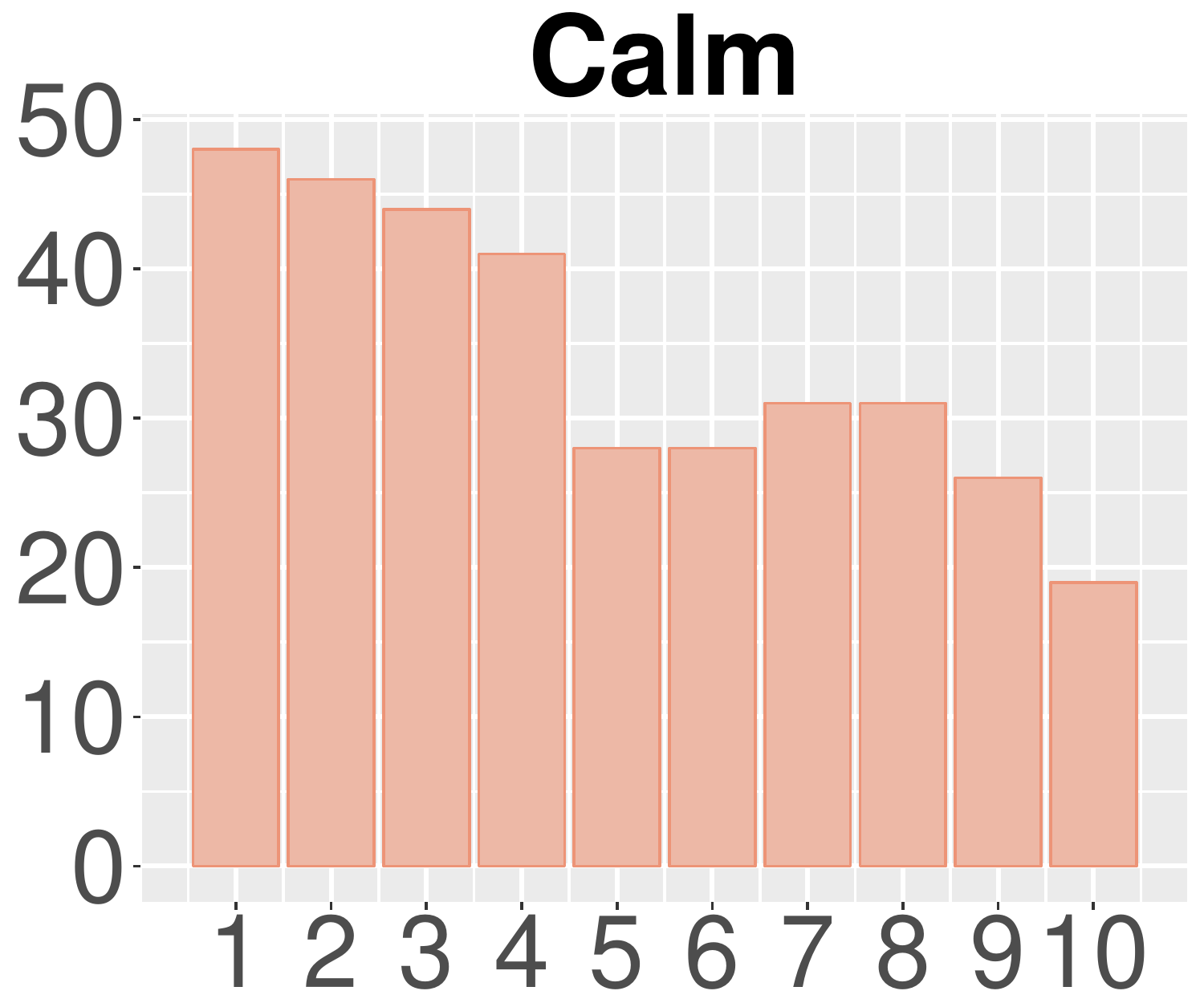} 
\includegraphics[width=0.24\columnwidth]{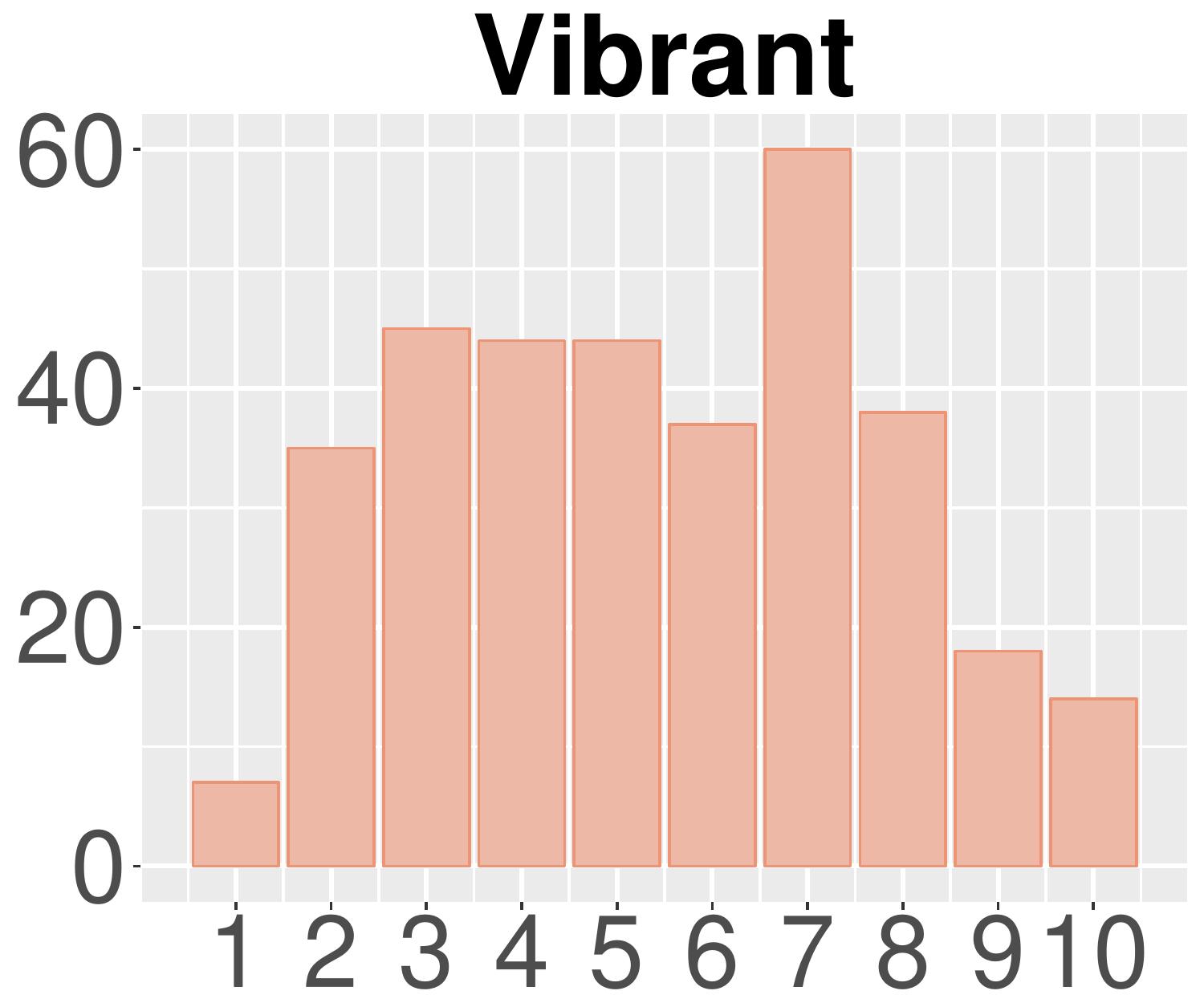}
\includegraphics[width=0.24\columnwidth]{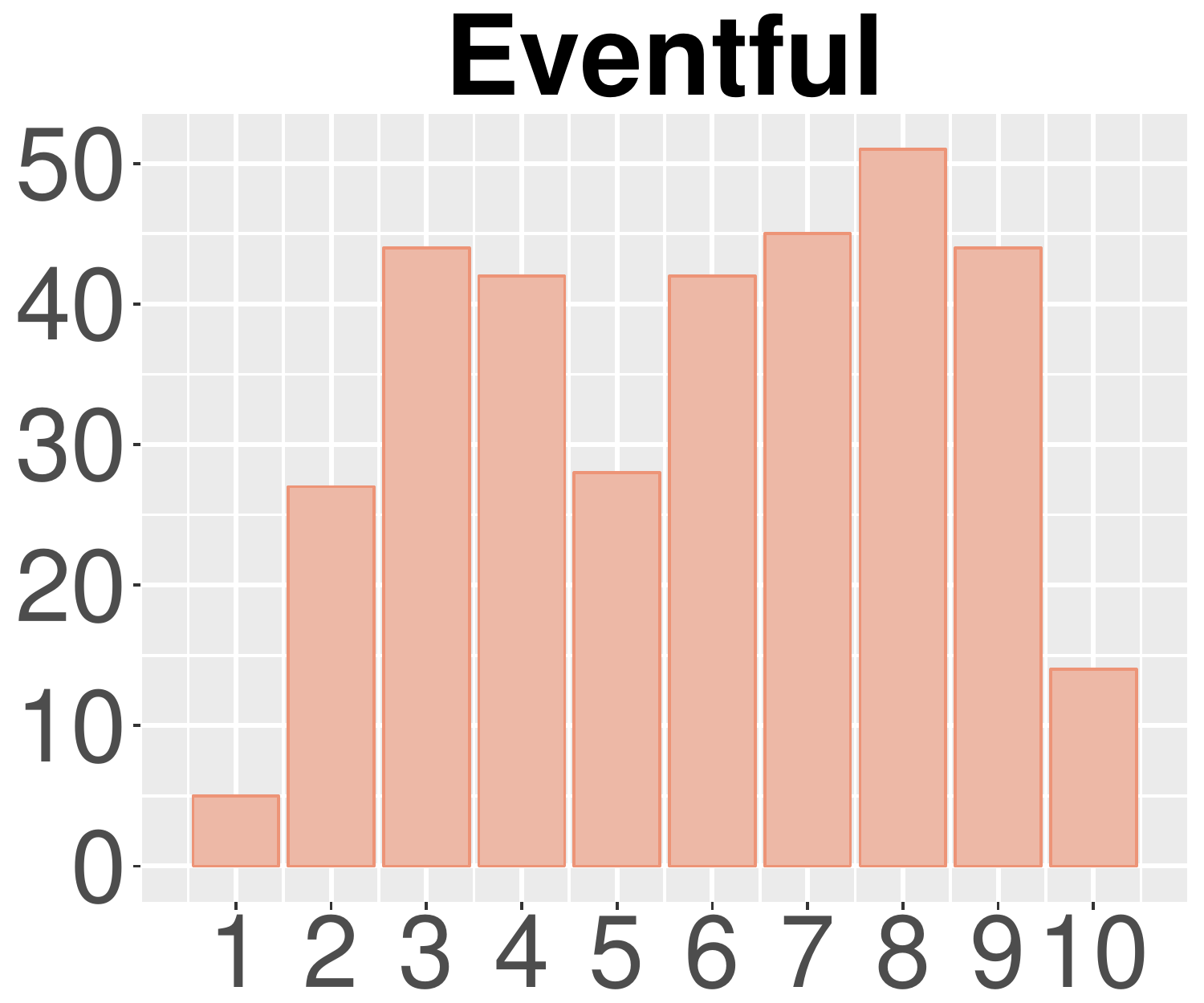}
\\
\includegraphics[width=0.24\columnwidth]{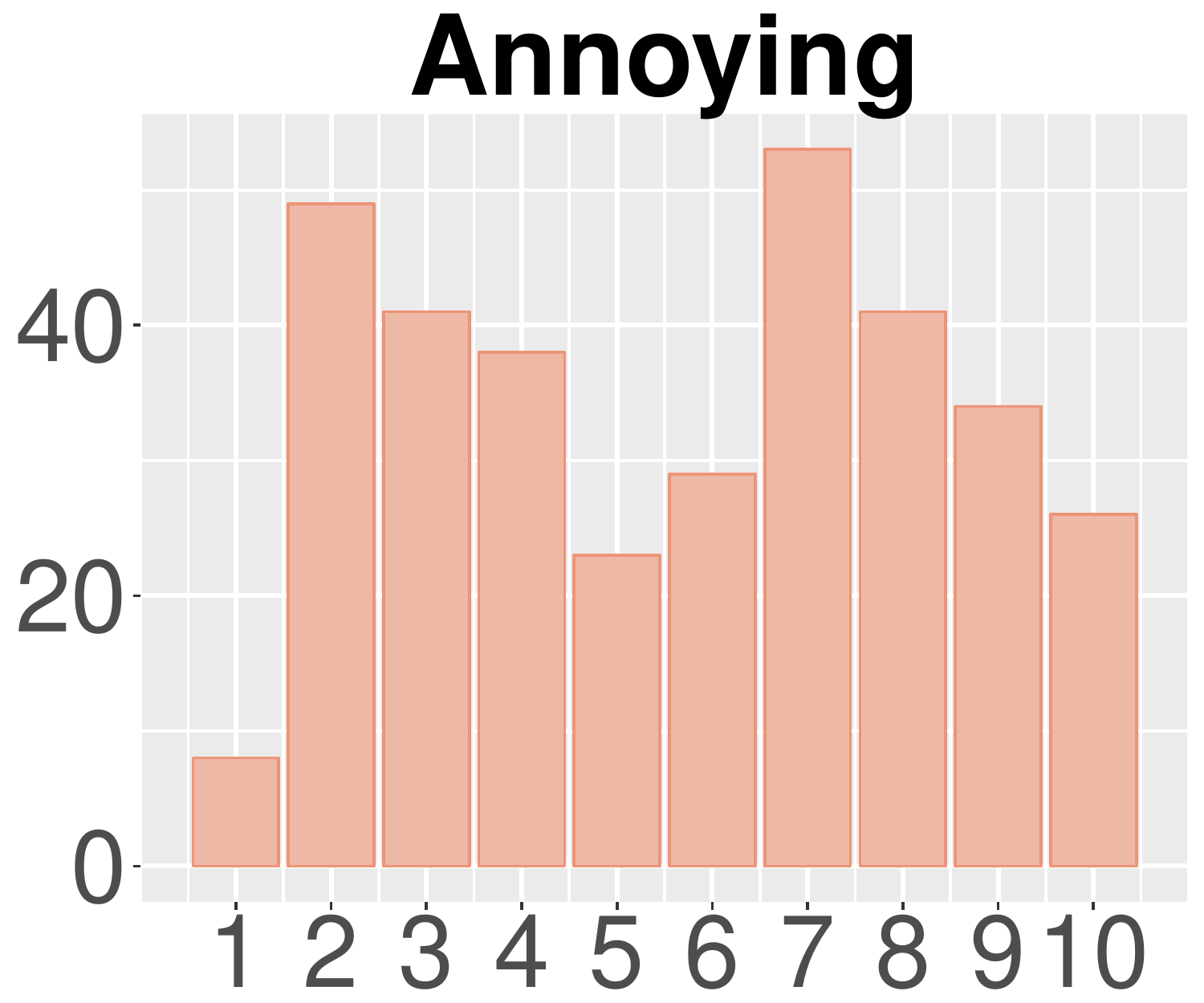} 
\includegraphics[width=0.24\columnwidth]{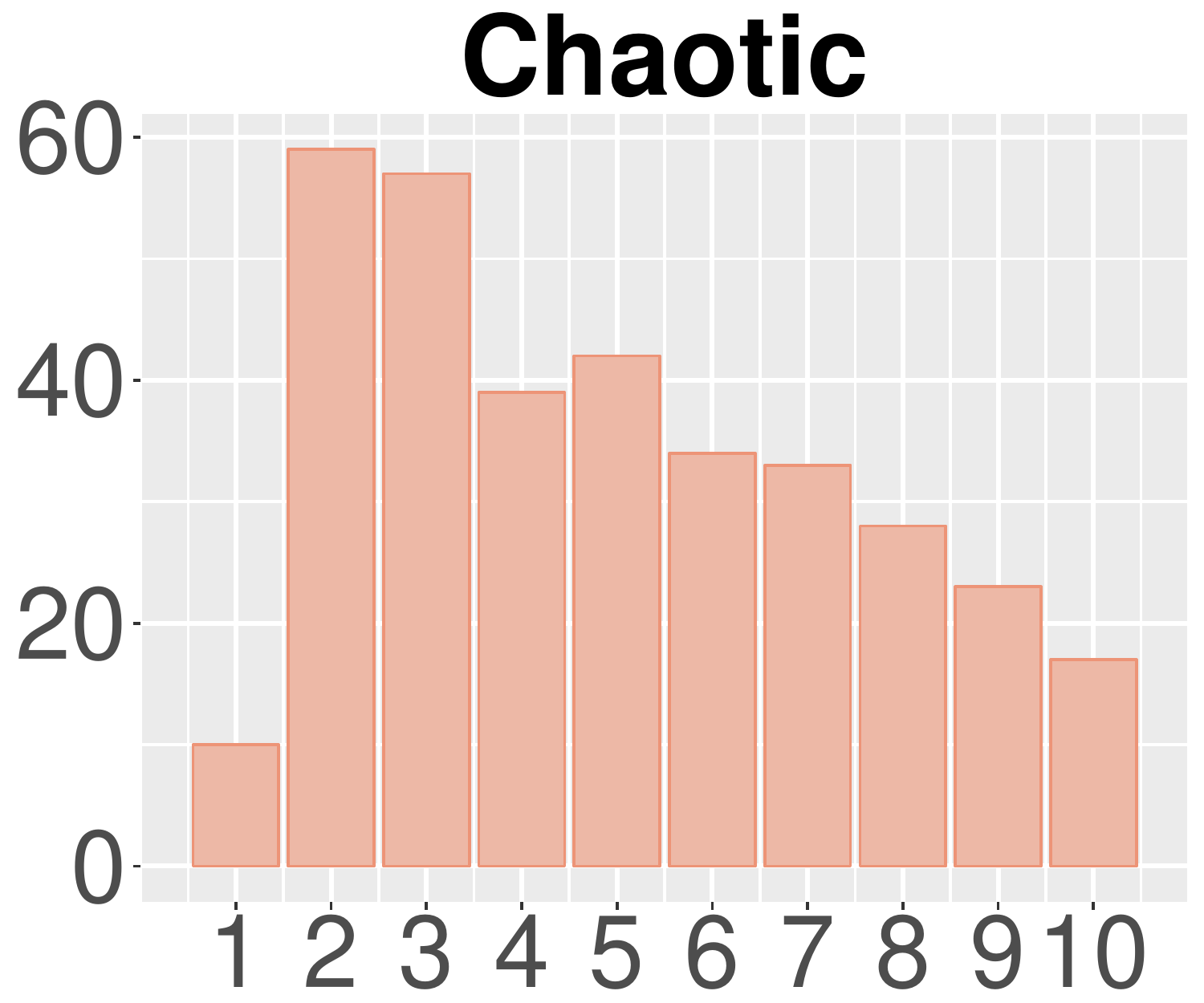} 
\includegraphics[width=0.24\columnwidth]{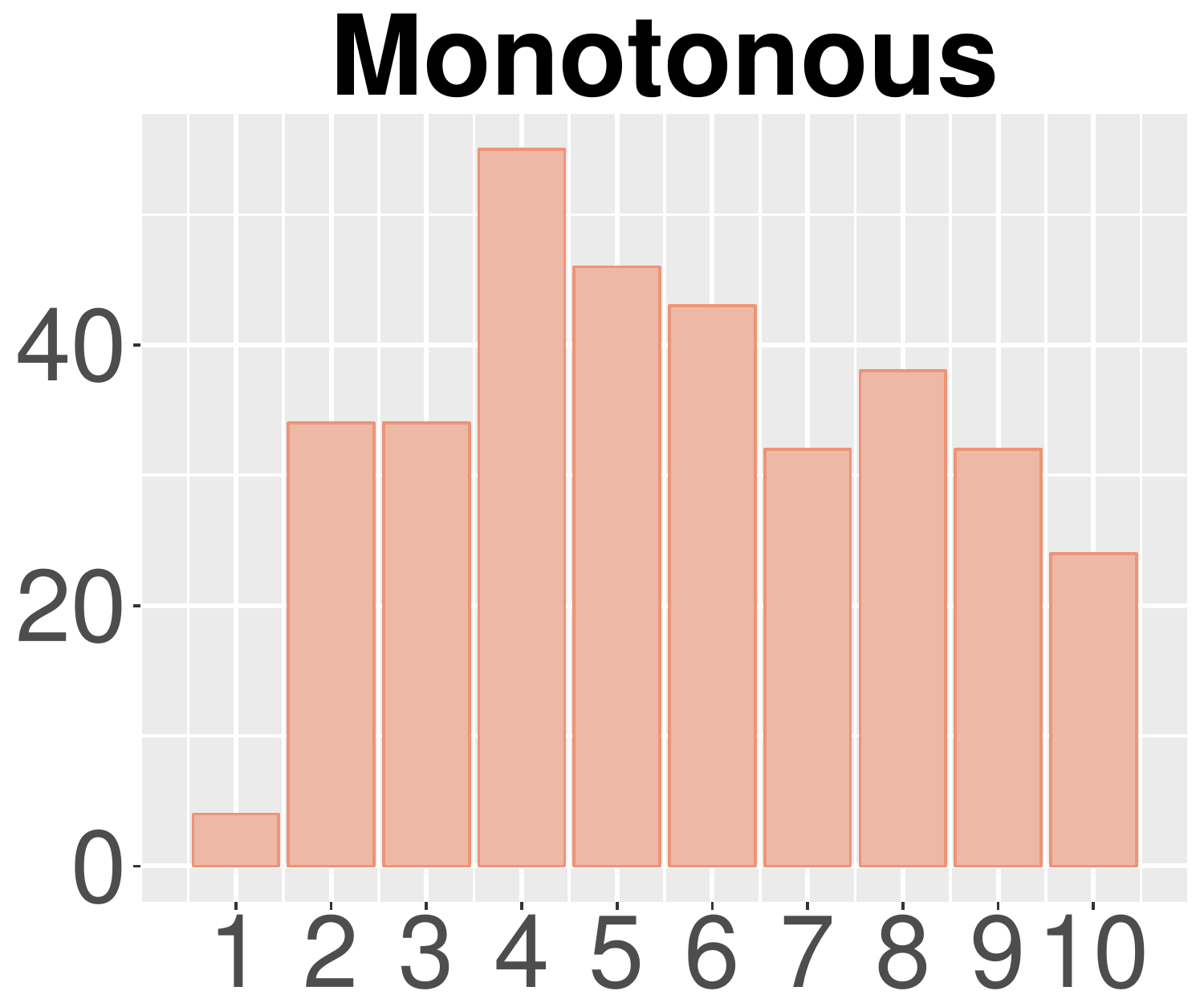}
\includegraphics[width=0.24\columnwidth]{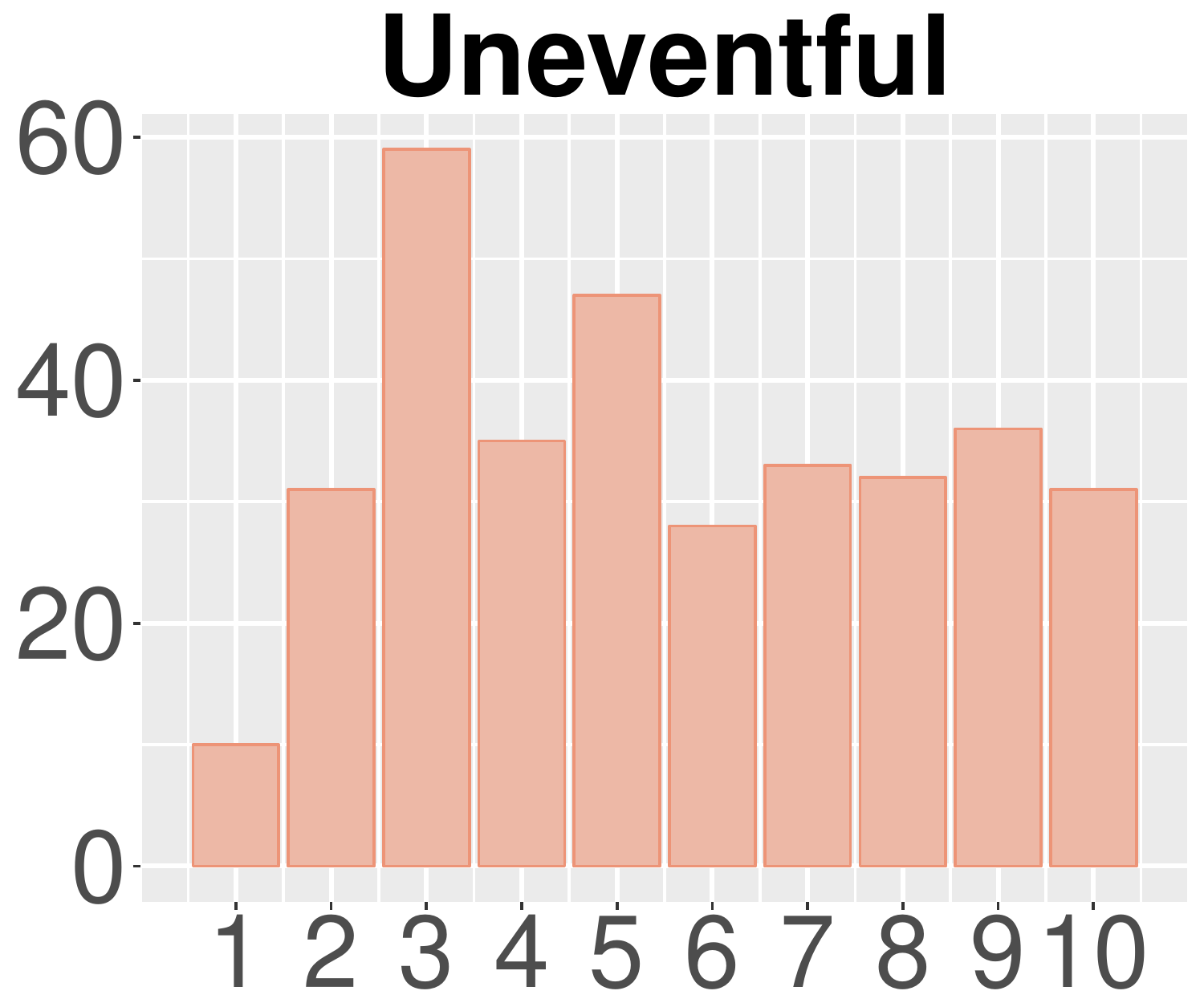}
\caption{Frequency distributions of the survey's perception scores (from 1 to 10) for each perception category. Most of the perceptions are scored  in  the full 1-to-10 range. }
\label{fig:perception_distr}
\end{figure}

\begin{figure}[tp]
\centering
\includegraphics[clip=true, width=.65\textwidth]{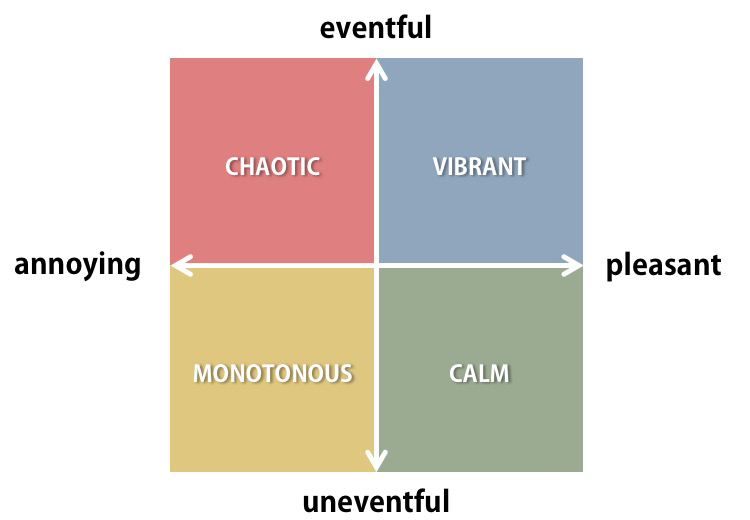}
\caption{Two principal components describing how study participants perceived urban sound. The combination of the first component ``uneventful \emph{vs.} eventful'' with the second component ``annoying \emph{vs.} pleasant'' results in four main ways of perceiving urban sounds: vibrant, calm, monotonous, and chaotic.}
\label{fig:quadrant}
\end{figure}

From our social media data, we knew the extent to which a potential source of sound was present on a street. If we knew how people usually perceived that source as well, we could have estimated how the street was likely to be perceived. 

One way of determining how people usually perceive sounds in the city context  is to run \textit{soundwalks}. These were introduced in the late 60s~\cite{southworth69sonic} and are still common  among acoustic researchers nowadays~\cite{semidor06listening,jeon13soundwalk}. Therefore, to determine people's perceptions, one of the authors conducted \textit{soundwalks} across   8 areas  in Brighton \& Hove (UK) and 11 areas in Sorrento (Italy) in April and October. They involved 37 participants (UK: 16 males, 5 females, $\mu_{age}$ = 38.6, $\delta_{age}$ = 11.5; Italy: 10 males, 6 females, $\mu_{age}$ = 34.7, $\delta_{age}$ = 7.1) with a variety of backgrounds (e.g., acousticians, architects, planning professionals, local authorities, and environmental officers). The experimenter led the participants along a pre-defined route and stopped at selected locations. At each of the locations, participants were asked to listen to the acoustic environment for two minutes and to complete a structured questionnaire (Table~\ref{tab:questionnaire}) inquiring about  sound sources' noticeability~\cite{axelsson09swedish}, soundscape attributes~\cite{axelsson09swedish}, overall soundscape quality~\cite{axelsson09swedish,liu13spatiotemporal} and soundscape appropriateness~\cite{axelsson15measure}. The questionnaire classified urban sounds into five categories (traffic, individuals, crowds, nature, other) as it is  typically done  in soundwalks~\cite{aletta15characterization,aletta15soundscape}, and the perceptions of such sounds into eight categories (pleasant, chaotic, vibrant, uneventful, calm, annoying, eventful, and monotonous, after Axelsson \emph{et al.}~\cite{axelsson10principal}'s work).

Those soundwalks resulted into 342 tuples, each of which represents a participant's report about sounds and perceptions at a given location. Each tuple had thirteen [1,10] values: five values reflecting the extent to which the five sound categories were reported to be present, and the other eight reflecting  the extent to which  the eight perceptions were reported. More technically, $sound_{k,c}$ is the score for sound category $c$  at tuple $k$, and $perception_{k,f}$ is the score for perception category $f$ at tuple $k$.  The frequency distributions of $sound_{k,c}$ (Figure~\ref{fig:perception_soundcategories_distr}) suggest that the participants experienced both streets with only a few sounds, and streets with many. Also, they rarely experienced crowds and came across traffic and, only at times, nature. Instead, the frequency distributions of $perception_{k,f}$ (Figure~\ref{fig:perception_distr})  suggest that the participants experienced streets with very diverse perceptual profiles, resulting in the use of the full [1,10] score range for all perceptions. 

\begin{figure}[tp]
\centering
\includegraphics[clip=true, width=.48\textwidth]{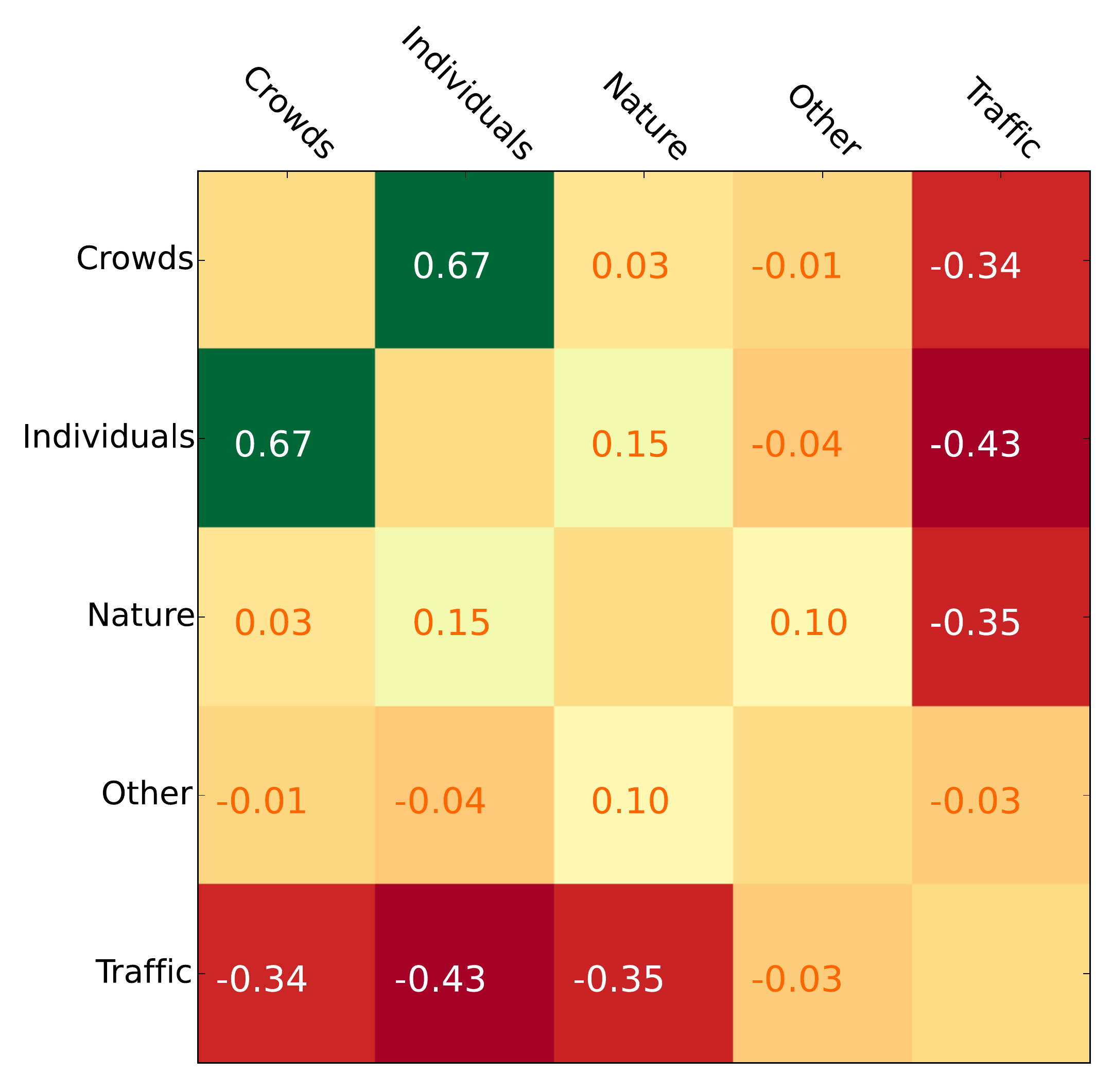}
\includegraphics[clip=true, width=.48\textwidth]{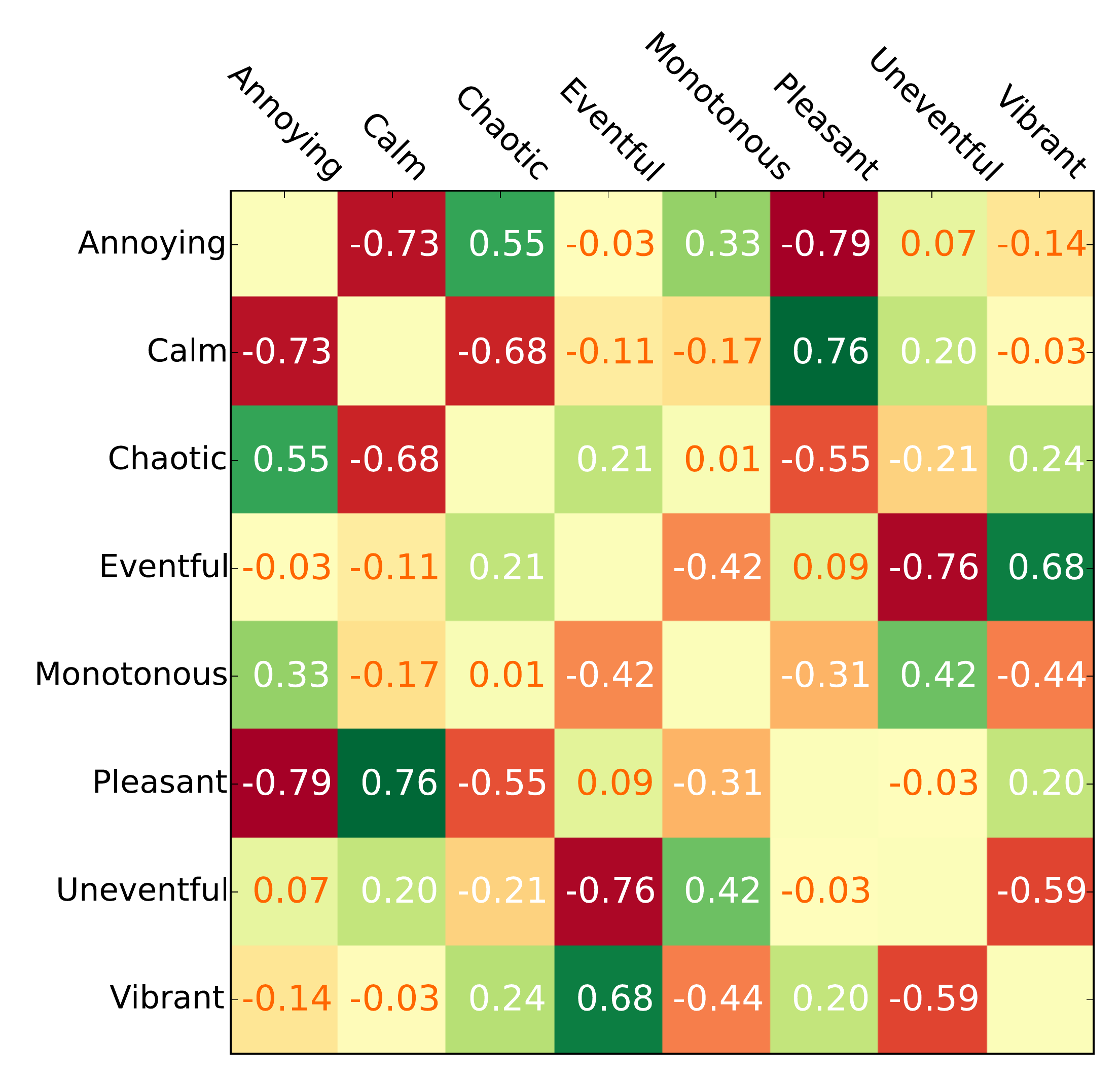}
\caption{Pairwise rank cross-correlations between the survey's sound scores  $sound_{k,c}$ (left) and its perception scores $perception_{k,e}$ (right).}
\label{fig:perceptions_cross_corr}
\end{figure}

To see which sounds participants tended to experience together, we computed the rank cross-correlation  $\rho_k(sound_{k,c_1},sound_{k,c_2})$ (left panel of Figure~\ref{fig:perceptions_cross_corr}). Amid crowds, the participants reported high score in the category `individuals'. These two sound categories --individuals and crowds-- had similar sound profiles so much so that the category `crowds' could be experimentally replaced by the category `individuals' in the specific instance of those soundwalks. Furthermore, as one would expect, the presence of traffic was associated with the absence of individuals, crowds, and nature.

To then see which perceptions participants tended to experience together, we computed the rank cross-correlation  $\rho(perception_{k,f_1}, perception_{k,f_2})$ (right panel of Figure~\ref{fig:perceptions_cross_corr}). Perceptions meant to have opposite meanings indeed resulted into negative correlations (pleasant \emph{vs.} annoying, eventful \emph{vs.} uneventful,  vibrant  \emph{vs.} monotonous, and calm  \emph{vs.} chaos). Interestingly, with their near-zero correlation, pleasantness and eventfulness were orthogonal - when a place was eventful, nothing could have been said about its pleasantness. 

\begin{figure*}[t!]
    \centering
    \begin{subfigure}[t]{0.5\textwidth}
        \centering
        \includegraphics[height=1.8in]{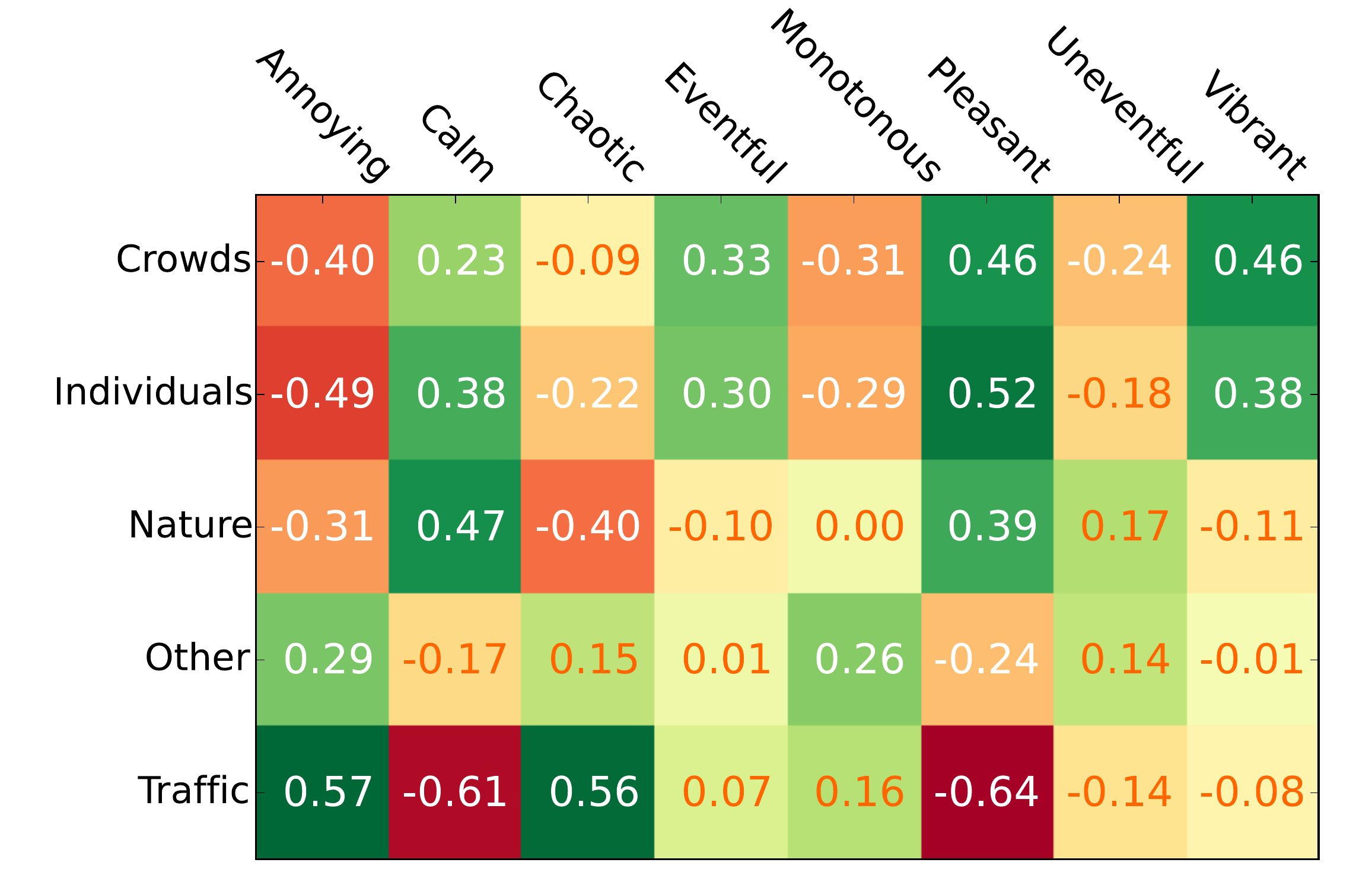}
        \caption{Correlations between the survey's sound scores $sound_{k,c}$ and its perception scores $perception_{k,e}$. Sounds of crowds, for example,  are perceived to be pleasant and vibrant but not annoying. }
    \end{subfigure}%
    ~ 
    \begin{subfigure}[t]{0.5\textwidth}
        \centering
        \includegraphics[height=1.8in]{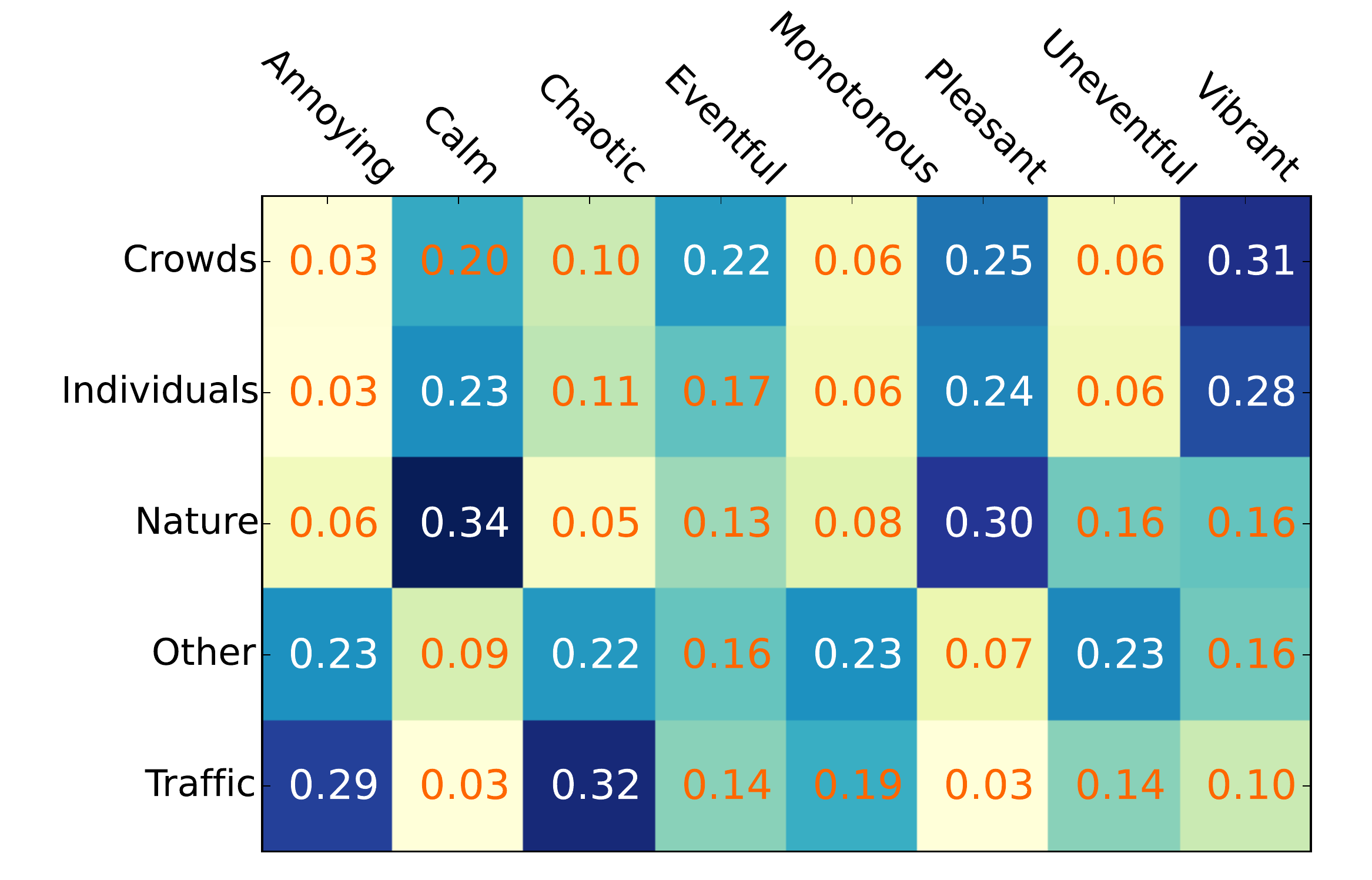}
        \caption{Probability $p(f|c)$  that perception $f$ was reported at a location with sound category $c$.}
    \end{subfigure}
    \caption{Relationship between sounds and perceptions in the soundwalk survey data.}
		\label{fig:perceptions_vs_sounds}
\end{figure*}

To see which sounds participants experienced together with which perception, we computed the rank correlation $\rho_k(sound_{k,c},perception_{k,f})$ (left panel of Figure~\ref{fig:perceptions_vs_sounds}). On average, vibrant areas tended to be associated with crowds, pleasant areas with individuals, calm areas with nature, and annoying and chaotic areas with traffic. In a similar way, Axelsson \emph{et al.} studied the principal components of their perceptual data~\cite{axelsson10principal} and found very similar results: they found that two components best explain most of the variability in the data (Figure~\ref{fig:quadrant}). 

Finally, to map how streets are likely to be perceived, we needed to estimate a street's \emph{expected} perception given the street's sound profile. The sound profiles came from our social media data, while the expected perception could have been computed from our soundwalks' data. We had already computed the correlations between sounds and perceptions (left panel of Figure~\ref{fig:perceptions_vs_sounds}). However, those correlations are not expected values (accounting for, e.g., whether a perception is frequent or rare) but they simply are strength measures. Therefore, we computed  the probability of perception $f$ given sound category $c$ as:
\begin{eqnarray}
& & p(f|c) = \frac{p(c|f) \cdot p(f)}{p(c)} 
\label{eq:conditional_probs}
\end{eqnarray}
To compute the composing probabilities, we needed to discretize our [1,10] values taken during the soundwalks, and did so by segmenting them into quartiles. We then computed:
\begin{eqnarray}
 & & p(c|f) = \frac{Q_4(c \wedge f)}{Q_4(f)}\\
 & & p(c) = \frac{Q_4(c)}{Q_4(c^*)} \quad;\quad  p(f) = \frac{Q_4(f)}{Q_4(f^*)}
\end{eqnarray}
where $Q_4(c)$ is the number of times the sound category $c$ occurred in the fourth quartile of its score; $Q_4(c^*)$ is the number of times \emph{any} sound occurred in its fourth quartile; and $Q_4(c \wedge f)$ is the number of times sound $c$ as well as perception $f$ occurred in their fourth quartiles.

The conclusions drawn from the resulting conditional probabilities (right panel in Figure~\ref{fig:perceptions_vs_sounds}) did not differ from those drawn from the previously shown sound-perception correlations (left panel). As opposed to the correlation values, none of the conditional probabilities were very high (all below $0.33$). This is because the conditional probabilities were estimated through the gathering of perceptual data in the wild\footnote{It has been shown that, in soundwalks, perception ratings are affected by not only sounds but also visual cues (e.g., greenery has been found to modulate soundscape `tranquillity' ratings~\cite{Watts11, Watts13}).} and, as such,  the mapping between perception and sound did not result in fully-fledged probability values. Those values are best interpreted not as raw values but as ranked values. For example, nature sounds were associated with calm only with a probability  0.34, yet calm is the strongest perception related to nature as it ranks first.

\begin{figure}[tp]
\centering
\includegraphics[clip=true, width=.90\textwidth]{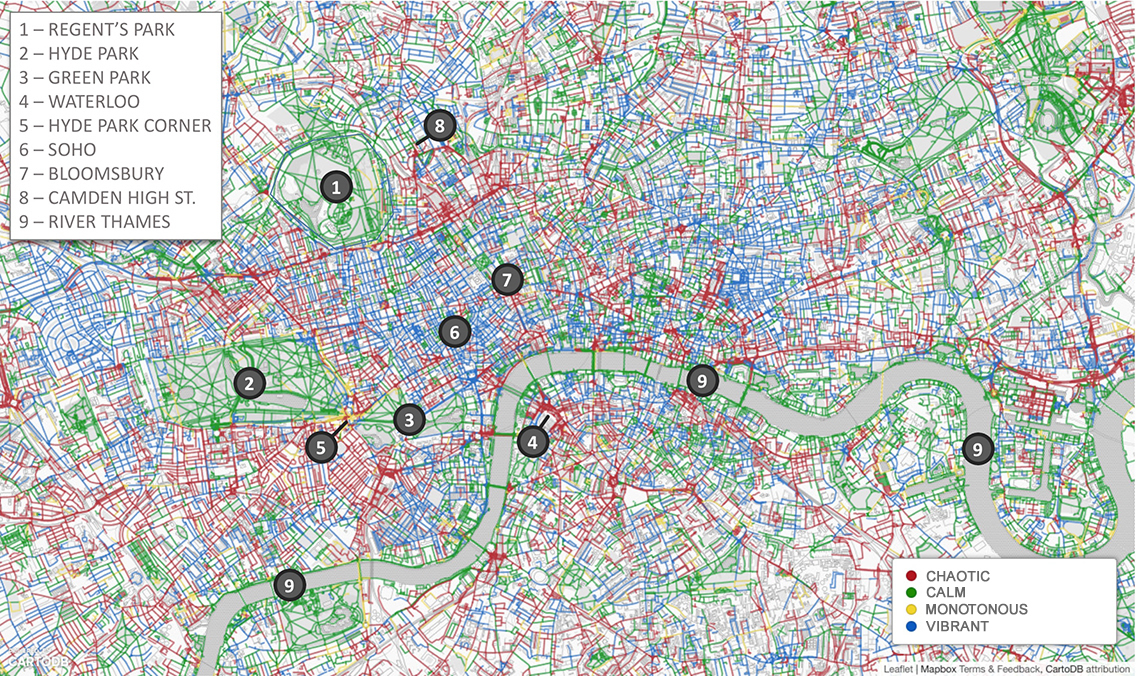}\\
\includegraphics[clip=true, width=.90\textwidth]{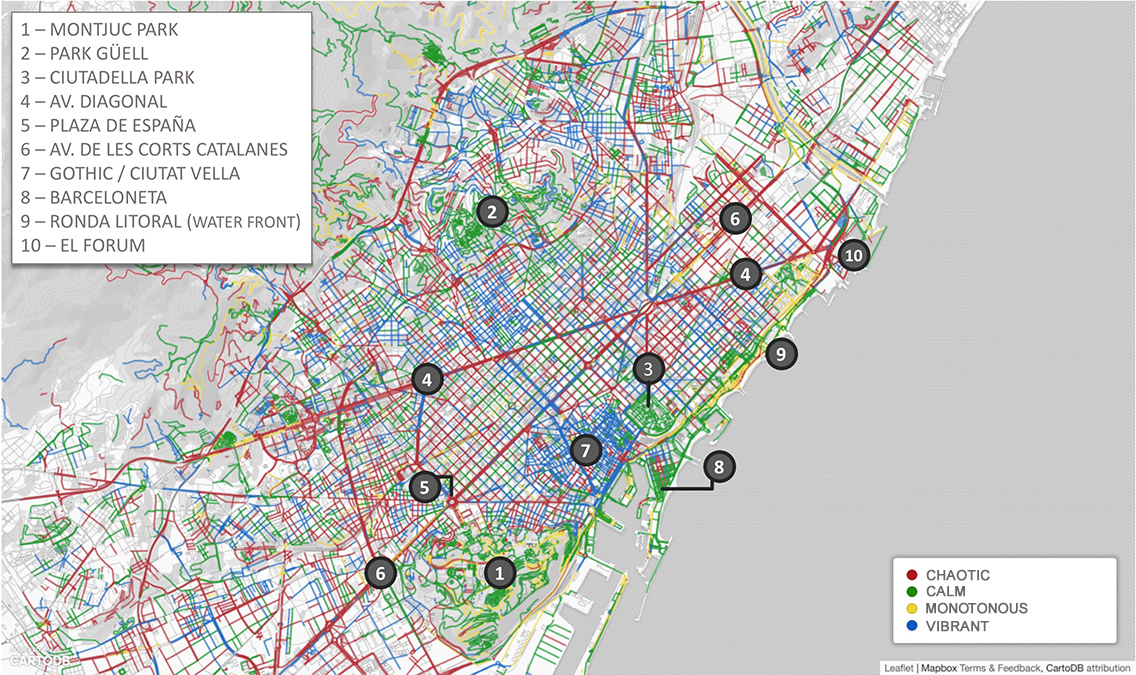}
\caption{Perceptual maps of London (top) and Barcelona (bottom). At each segment, the perception $f$ with the highest probability was reported (i.e., with the highest $p_j(f)$). In London, calm sounds were found in  Regent’s Park(1); Hyde Park(2);  Green Park(3); and all around the  River Thames(9). By contrast,  chaotic sounds were around Waterloo station(4) and Hyde Park corner (5). Vibrant sounds were found in Soho(6), Bloomsbury(7), and Camden High Street(8). In Barcelona,  calm sounds  were found in  Montjuic Park(1); Park Guell(2); and  Ciutadella Park(3), and on the beach of Barceloneta(8). By contrast, on the beach in front of Ronda Litoral(9), we found monotonous sounds. Chaotic sounds were found on the main road of  Avinguda Diagonal(4), on Plaza de Espana(5), and on Avinguda De Les Corts Catalanes(6). Vibrant sounds were found in the historical center called Gothic/Ciutat Vella(7), and a bit in the open-air arena of El Forum(10), which was also characterized by chaotic sounds.}
\label{fig:perception_map}
\end{figure}

The advantage of conditional probabilities over correlations is that they offer principled numbers that are properly normalized and could be readily used in future studies.  They could be used, for example, to draw an aesthetics map, a map that reflects the emotional qualities of sounds. In the maps of Figure~\ref{fig:perception_map}, we associated each segment with the color corresponding to the perception with the highest value of $p_j(f)= \sum_c p(f|c) \cdot p_j(c)$, where $p_j(c)=sound_{j,c}$, which is the fraction of tags at segment $j$ that matched sound category $c$. $p_j(f)$ is effectively the probability  that perception $f$ is associated with street segment $j$, and the strongest $f$ is associated with $j$. By mapping the probabilities of sound perceptions in London (top panel of Figure~\ref{fig:perception_map}) and Barcelona (bottom panel), we observed that  trafficked roads were chaotic, while walkable parts of the city were exciting. More interestingly, in the soundscape literature, monotonous areas have not necessarily been considered pleasant (they fall into the annoying quadrant of Figure~\ref{fig:quadrant}), yet the beaches of Barcelona were monotonous (and rightly so), but might have been pleasant as well. 

\begin{table}[tp]
\footnotesize
\begin{tabularx}{390pt}{XXX}
\multicolumn{1}{c}{\textbf{Question}} & \multicolumn{1}{c}{\textbf{Items}} & \multicolumn{1}{c}{\textbf{Scale extremes (1-10)}} \\
\hline
To what extent do you presently hear the following five types of sounds? & Traffic noise (e.g., cars, trains, planes), sounds of individuals (e.g., conversation, laughter, children at play), crowds of people (e.g., passers, sports event, festival), natural sounds (e.g., singing birds, flowing water, wind in the vegetation), other noise (e.g. sirens, construction, industry) & [Do not hear at all,  \ldots , Dominates completely]\\ 
\hline
Overall, how would you describe the present surrounding sound environment? & \multicolumn{1}{c}{---} & [Very bad, \ldots, Very good] \\ 
\hline
Overall, to what extent is the present surrounding sound environment appropriate to the present place? & \multicolumn{1}{c}{---} & [Not at all, \ldots, Perfectly] \\ 
\hline
For each of the 8 scales below, to what extent do you agree or disagree that the present surrounding sound  environment is... & pleasant, chaotic, vibrant, uneventful, calm, annoying, eventful, monotonous & [Strongly disagree, \ldots, Strongly agree]\\ 
\hline
\end{tabularx}
\caption{The questionnaire used during the soundwalk. For each question, participants could express their preference on a 10-point ordinal scale.}
\label{tab:questionnaire}
\end{table}

\section{Discussion}\label{sec:limitations}

\begin{figure}[tp]
\centering
\includegraphics[clip=true, width=.32\textwidth]{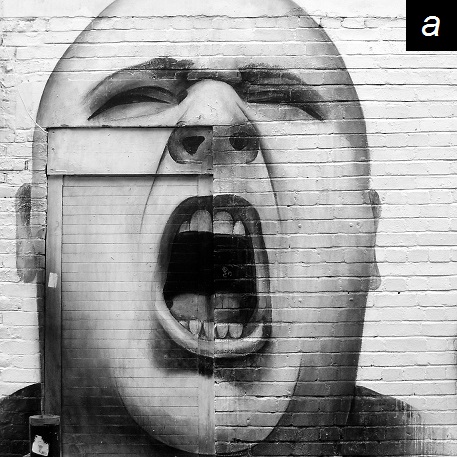}
\includegraphics[clip=true, width=.32\textwidth]{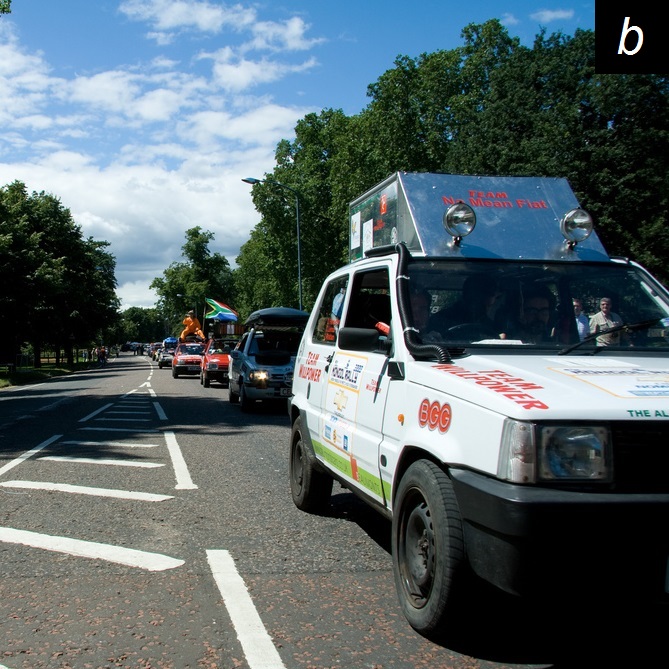}
\includegraphics[clip=true, width=.32\textwidth]{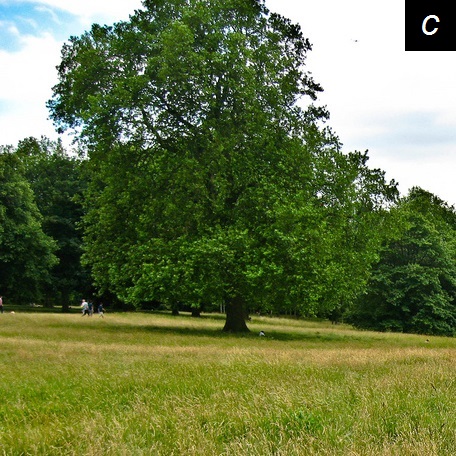}
\caption{Examples of  ambiguously tagged pictures. \emph{(a)} street art in Brick Lane tagged with the term ``screaming'', and the same location Carriage Drive with Hyde Park tagged with opposing terms related to \emph{(b)} traffic sounds and  \emph{(c)} nature sounds.}
\label{fig:examples}
\end{figure}

A project called SmellyMaps mapped urban smellscapes from social media~\cite{quercia15smelly}, and this work - called ChattyMaps - has three main similarities with it. First, the taxonomy of sound and that of smell were both created using community detection algorithms, and both closely resembled categorizations widely used by researchers and practitioners in the corresponding fields. Second, the ways that social media data was mapped onto streets (e.g., buffering of segments, use of long/lat coordinates on the pictures) is the same.  Third, in both works, the validation was done with official data  (i.e., with air quality data and noise pollution data). However,  the two works differ as well, and they do so in three main ways. First, as opposed to SmellyMaps, ChattyMaps studied a variety of urban layers: not only the urban sound layer but also the emotional, perceptual, and sound diversity layers. Second, smell words were derived from smellwalks (as no other source was available), while sound words were derived from the online platform of Freesound.  Third, since SmellyMaps showed that picture tags were more effective than tweets in capturing geographic-salient information,  ChattyMaps entirely relied on Flickr tags. 

Our approach comes with a few limitations, mainly because of data biases. The urban soundscape is multifaceted: the sounds we hear and the way we perceive them change considerably with small variations of, for example, space (e.g., simply turning a corner) and time (e.g., day \emph{vs.} night). By contrast, social media data has limited resolution and coverage, and that results into false positives. At times, sound tags do not reflect real sounds  because of  either misannotations or the figurative use of tags (picture Figure~\ref{fig:examples}(\textit{a})). Fortunately, those cases occur rarely. By manually inspecting 100 photos with sound tags, no false positive was found:  87 pictures were correctly tagged, and 13 referred to sounds that were plausible yet hard to ascertain. 

Even when tags refer to sounds likely present in an area, they might do so partially. For example, the tags on the picture of Figure~\ref{fig:examples}(\textit{b}) consisted of traffic terms (rightly) but not of nature terms, and that was a partial view of that street's soundscape.  This risk shrinks as the number of sound tags for the segment increases. Indeed, let us stick with the same example: Figure~\ref{fig:examples}($c$) was taken few meters away from ($b$), and its tags consisted of nature terms. 

To partly mitigate noise at boundary regions, we did two things. First, as described in Section~\ref{sec:method}, we added a buffer of 22.5 meters around each segment's bounding box. This has been commonly done  in previous work dealing with geo-referenced digital content~\cite{quercia15walk,quercia15smelly}. It is hard to measure automatically how many tags are needed to get high confidence sound profiles, but we estimated it to be around 20-25 tags (Figure~\ref{fig:correlation_sound_pollution}), if official air quality data is used for validation. 

Second, we associated sound \emph{distributions} (and not individual sounds) with street segments. The 6-dimensional sound vector was normalized in [0,1] to have a probabilistic interpretation. In Figures~\ref{fig:examples}(\textit{b}) and (\textit{c}), nature sounds were predominant, yet  traffic-related sounds varied from $20\%$ to $2\%$ depending on the different parts of that street. 

\begin{figure}[tp]
\centering
\includegraphics[clip=true, width=.99\textwidth]{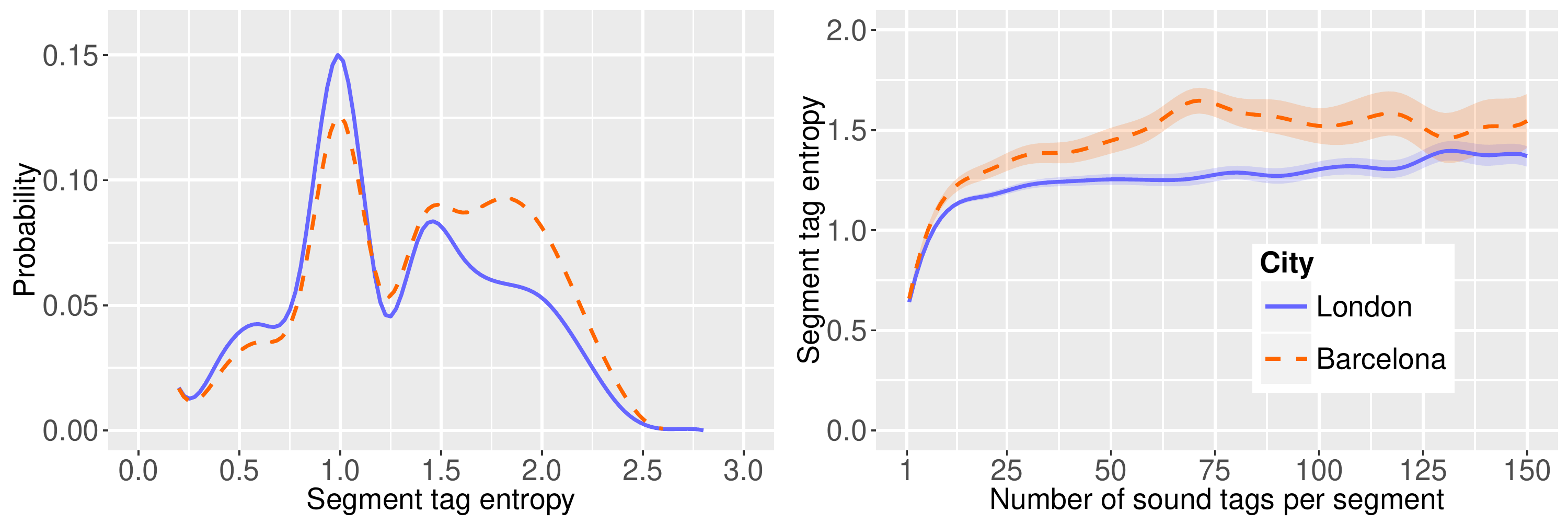}
\caption{Diversity (entropy) of sound tags. Frequency distribution (left), and how the diversity varies with the number of tags per street segment (right). Segments with zero diversity ($28\%$ in Barcelona, $35\%$ in London) were excluded.}
\label{fig:entropy}
\end{figure}

\begin{figure}[tp]
\centering
\includegraphics[clip=true, width=.90\textwidth]{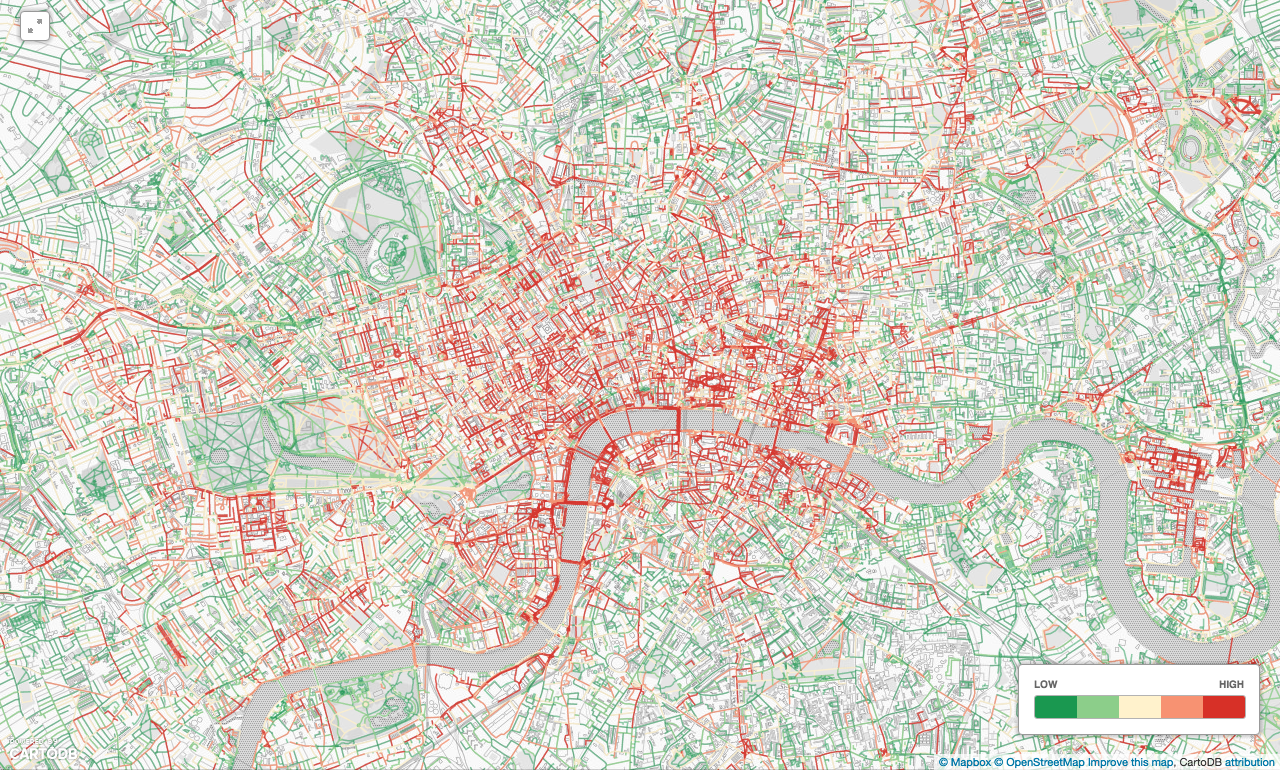}\\
\includegraphics[clip=true, width=.90\textwidth]{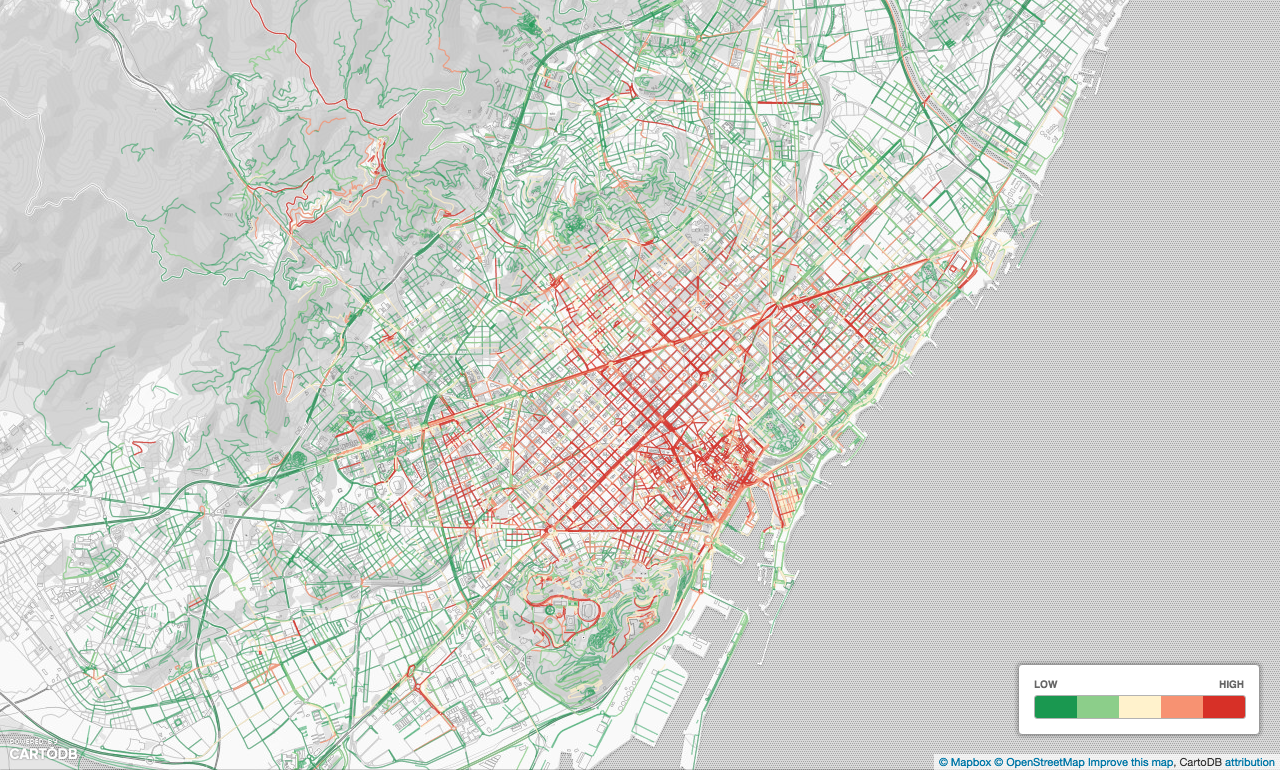}
\caption{Maps of the diversity of sound tags for each street segment in London (top) and Barcelona (bottom). Only segments with 5 or more tags are displayed.}
\label{fig:entropy_map}
\end{figure}

More generally, to have a more comprehensive view of this phenomenon, we determined each segment's sound diversity by computing the Shannon index:
\begin{eqnarray}
& & diversity_j= - \sum_c sound_{j,c} \cdot ln(sound_{j,c})
\label{eq:shannon}
\end{eqnarray}
where $sound_{j,c} $ is the fraction of tags at segment $j$ that matched sound category $c$. After removing zero diversity values (often associated with segments having only one tag, which made 28\% of segments in Barcelona, and 35\% segments in London), we saw that the frequency distribution of diversity (Figure~\ref{fig:entropy},~left) had two peaks in 1 (for both cities) and in 1.5 for London and in 2.0 for Barcelona. Then, by mapping those values (Figure~\ref{fig:entropy_map}), we saw that the values close to the first peak were associated with parks and suburbs, and those close to the second peak (and higher) were associated with the central parts of the two cities. Furthermore, the diversity did not depend on the number of tags per segment and became stable for segments with at least 10 tags (Figure~\ref{fig:entropy},~right). 


\section{Conclusion}\label{sec:conclusions}

We showed that social media data makes it possible to effectively and cheaply track urban sounds at scale.  Such a tracking was effective because the resulting sounds were  geographically sorted across street types  in expected ways, and they matched noise pollution levels. The tracking was also cheap because it did not require the creation of any additional service or infrastructure. Finally, it worked at the scale of an entire city, and that is important, not least because, before our work, there had  been \textit{``nothing in sonography corresponding to the instantaneous impression which photography can create \ldots The microphone samples details and gives the close-up but nothing corresponding to aerial photography.''}~\cite{schafer77soundscape}

However, whereas landscapes can be static, soundscapes are dynamic~\cite{blesser2009spaces}. Their perceptions are affected by  demography (e.g., personal sensitivity to noise, age), context (e.g., city layout), and time (e.g., day \emph{vs.} night, weekdays  \emph{vs.} weekends). Future studies could partly address those issues by collecting additional data and by comparing models of urban sounds generated from social media with those generated from GIS techniques.  

Nonetheless,  no matter what data one has, fully capturing soundscapes might well be impossible. Our work has focused on identifying potential sonic events. To use a food metaphor, those events are the raw ingredients, then the aural architecture (which comes with the acoustic properties of trees, buildings, streets) is the cooking style, and the soundscape is the dish~\cite{blesser2009spaces}. 

To unite hitherto isolated studies in a new synergy, in the future, we will conduct a  comprehensive multi-sensory research of cities, one in which visual~\cite{quercia14shortest, quercia2014aesthetic}, olfactory~\cite{quercia15smelly}, and sound perceptions are explored together. 

The ultimate goal of this work is to empower city managers and researchers to \textit{``find solutions for an ecologically balanced soundscape where the relationship between the human community and its sonic environment is in harmony''}, as Schafer famously (and prophetically) remarked in the late 70s~\cite{schafer77soundscape}.

\section*{Acknowledgments} 
We thank the Barcelona City Council for making the noise pollution data available. F. Aletta received funding through the People Programme (Marie Curie Actions) of the European Union's $7^{th}$ Framework Programme FP7/2007-2013 under REA grant agreement \textit{n} 290110, SONORUS ``Urban Sound Planner''.

\bibliographystyle{abbrv}
\bibliography{bibs}

\end{document}